\documentclass[twocolumn]{emulateapj}
\usepackage{epsf}
\usepackage{amsmath}
\usepackage{multirow}
\usepackage{rotating}
\usepackage{subfigure}
\usepackage{multirow}
\usepackage{color}
\begin{document}

\slugcomment{Accepted to ApJ on 14 May 2013}
\shortauthors{A. L. King et al.}
\shorttitle{Influence of Spin in Compact Objects}

\title{What is on Tap? The Role of Spin in Compact Objects and Relativistic Jets}

\author{Ashley~L.~King\altaffilmark{1},
         Jon~M.~Miller\altaffilmark{1},
	Kayhan~G\"ultekin\altaffilmark{1},
	Dominic~J.~Walton\altaffilmark{2},
         Andrew~C.~Fabian\altaffilmark{3},
         Christopher~S.~Reynolds\altaffilmark{4},
	Paul~Nandra\altaffilmark{5}}
       
\altaffiltext{1}{Department of Astronomy, University of Michigan, 500
Church Street, Ann Arbor, MI 48109-1042, ashking@umich.edu}
\altaffiltext{2}{Space Radiation Laboratory, California Institute of Technology, Pasadena, CA 91125, USA}
\altaffiltext{3}{Institute of Astronomy, University of Cambridge, Madingley Road, Cambridge, CB3 OHA, UK}
\altaffiltext{4}{Department of Astronomy, University of Maryland, College Park, MD 20742, USA}
\altaffiltext{5}{Max Planck Institute for Extraterrestrial Physics, 85741 Garching, Germany}

\label{firstpage}

\begin{abstract}
We examine the role of spin in launching jets from compact objects across the mass scale. Our work includes three different Seyfert samples with a total of 37 unique Seyferts, as well as 11 stellar-mass black holes, and 13 neutron stars. We find that when the Seyfert reflection lines are modeled with simple Gaussian line features (a crude proxy for inner disk radius and therefore spin), only a slight inverse correlation is found between the Doppler-corrected radio luminosity at 5 GHz (a proxy for jet power) and line width. When the Seyfert reflection features are fit with more relativistically-blurred disk reflection models that measure spin, there is a tentative positive correlation between the Doppler-corrected radio luminosity and the spin measurement. Further, when we include stellar-mass black holes in the sample, to examine the effects across the mass scale, we find a slightly stronger correlation with radio luminosity per unit mass and spin, at a marginal significance (2.3$\sigma$ confidence level). Finally, when we include neutron stars, in order to probe lower spin values, we find a positive correlation (3.3$\sigma$ confidence level) between radio luminosity per unit mass and spin.  Although tentative, these results suggest that spin may have a role in determining the jet luminosity. In addition, we find a slightly more significant correlation (4.4$\sigma$ and 4.1$\sigma$ confidence level, respectively) between radio luminosity per Bolometric luminosity and spin, as well as radio luminosity corrected for the fundamental plane (i.e, $\log  (\nu L_R/L_{Bol}^{0.67}/M_{BH}^{0.78})$) and spin, using our entire sample of black holes and neutrons stars. Again, although tentative, these relations point to the possibility that the mass accretion rate, i.e. Bolometric luminosity, is also important in determining the jet luminosity, in addition to spin. Our analysis suggests that mass accretion rate and disk or coronal magnetic field strength may be the ``throttle" in these compact systems, to which the Eddington limit and spin may set the maximum jet luminosity that can be achieved. 
\end{abstract}

\section{Introduction}

Observationally, the large-scale impact of relativistic jets from black holes is well constrained via radio and X-ray observations. Classical double lobed structures are seen on kpc to Mpc scales as the charged particles emit via synchrotron emission and in shocks along the jet and in the impact lobes \citep[e.g.,][]{Bridle84}. Jets may even be responsible for heating of the intragalactic and intracluster medium \citep[e.g.,][and references therein]{Fabian12}. However, on AU to pc scales, observations fail to resolve the launching regions \citep[except in one ideal scenario of M87,][]{Doeleman12}. Therefore, we must rely on other methods to infer the launching characteristics of these immense and powerful structures. 

Our aim is to determine the vital processes that regulate the launching and collimation of these powerful outflows. Many theories predict that jets can tap the angular momentum, or ``spin'' ($a=cJ/GM^2$, $-1<a<1$), of the compact object to transfer immense amounts of energy from the compact object to the outflows. \cite{Blandford77} describe this process by threading magnetic field lines through the ergosphere of the black hole to spin down the black hole and transfer energy to the jet. At low spin, and for a parabolic magnetic field geometry, the jet emission scales as, $L_{Jet} \propto a^2 B^2 M^2_{BH}$, where $a$ is the spin, $B$ is the magnetic field, and $M_{BH}$ is the mass of the black hole. If the accretion disk is efficiently radiating, this luminosity also scales with the accretion disk luminosity as $L_{Jet} \propto a^2 L_{Disk}$. 

As jet luminosity has been used a proxy for spin, it is important to understand the actual role of spin in samples where we can measure both the spin and the jet luminosity. In addition, spin has been used to explain the dichotomy between radio-loud and radio-quiet galaxies \citep[e.g.,][]{Wilson95}. Apart from a dramatic difference in radio luminosities, these galaxies appear to have similar mass and mass accretion rates. Determining if spin has a role to play in this division is therefore very pertinent.

It is imperative to test these models via the most direct observations as possible. In particular, using the X-ray band in conjunction with broad-band radio observations allows for the most direct detection of the key parameters in these theoretical jet models. The X-ray band covers the typical band where relativistic reflection features from the inner accretion disk are found. The spin of the black hole has a direct effect on the line width and shape of these features. Therefore, accurately modeling these features is key to determining the spin \citep[e.g.,][]{Miller07}. In addition, the X-ray continuum is thought to be a proxy for the mass accretion rate of these systems, which is also likely to influence the total amount of jet power expelled from the system. The corresponding compact radio emission associated with the central engine is often used as a tracer for the jet power via synchrotron emission. 

Although all black holes are likely capable of producing such outflows, particular classes of black holes make the study of both jets and the inner accretion disk more favorable than others. Seyfert galaxies harbor supermassive black holes that are accreting at $~10^{-3}-10^{-2} L_{Edd}$. They are relatively nearby, with little obscuration, and therefore bright galaxies with high count rates. And even though Seyferts accrete at high Eddington rates, they still have compact radio emission with flat spectral indices that indicate synchrotron emission, typical of a jet \citep[e.g.,][]{King11}. Further, there are a larger number of Seyferts that have been well studied giving the highest number of sources in a particular black hole sample.

In this paper, we aim to characterize a number of different Seyfert samples, in order to determine if spin is a contributing factor to jet power. We will compare the line widths, followed by more physical spin measurements of Seyferts to several different jet power proxies involving the radio luminosity  and mass of each Seyfert. Finally, we will compare these Seyferts to their stellar-mass analogs, both black hole binaries and neutron stars, in an aim to understand how jet production and spin changes across the mass scale. By also comparing black holes and neutron stars, we can examine the global trends across compact objects and not just in black holes.

\section{Methods}

We have compiled a sample of several X-ray surveys of Seyfert galaxies from \cite{Nandra07}, \cite{Patrick12}, and \cite{Walton12}. We begin with the studies that use simple Gaussians lines to characterize the X-ray excess between 6.4--6.97 keV. We require:
\begin{enumerate}
\item a detection of the Fe K$\alpha$ lines at greater than a 3 $\sigma$ confidence level, defined as improving the fit statistics when adding a Gaussian line, 
\item that the Gaussian line width be inconsistent with 0 keV, so as to exclude the narrow features which likely arise further out in the disk, broad line region or torus, 
\item a complimentary radio observation of the source, i.e both detections and upper limits. 
\end{enumerate}
The Gaussian line width may be proportional to the radial extent of the disk. As the reflecting surface gets closer to the black hole, the Doppler broadening will increase the line width. Our later samples also include complex, self-consistent reflection models of the Fe lines as well as the soft excess seen in Seyferts \citep{Walton12}. These models not only measure the spin parameter ($a=cJ/GM^2$, $-1<a<1$), which is dependent on the radial position of the inner edge of the accretion disk, but also the inclination and emissivity of the emitting regions. Again, we require a statistically significant detection (3$\sigma$) of the Fe line in independent fits as well as complimentary radio observations. 

The radio luminosity is taken to be the luminosity at 5 GHz and corrected for Doppler boosting. We use the following relation to correct for Doppler boosting, adopted from \cite{Mirabel99},
\begin{equation}
\frac{S_{obs}}{S_{emit}} = \delta^{k-\alpha}
\end{equation}
where $S_{obs}$ is the observed flux density, $S_{emit}$ is the emitted flux density, $\delta = (\Gamma(1-\beta\cos\theta))^{-1}$, $\theta$ is the inclination angle, $k$ describes whether the emission is discrete ($k=3$) or continuous ($k=2$), and $\alpha$ is radio spectral index. For our work we assume that $\Gamma$=5 and $k=3$ for all sources. In addition, when the spectral index was unknown, we use $\alpha = -0.7$ for our Seyferts and $\alpha = -0.3$ for BHB, which is consistent with the median of measured spectral indices in each of these samples. 

We also include a sample of stellar-mass black holes with reflection line spin measurements for comparison across the mass scale. This sample is made up of black hole binaries (BHB) with both Fe lines that have reflection modeling as well as detected radio emission. As the timescale for jet production is much shorter in BHB than in supermassive black holes (SMBH), we chose the highest peak flux density listed in the literature to use in our analysis. Finally, we include a sample of neutron stars. These compact objects have observationally, systematically lower spins than either the supermassive or stellar-mass black holes, thus extending our range in the spin parameter. We note that we do not correct for Doppler boosting in this sample because jets from neutron stars are slower than their black hole counterparts. However, results should still be considered with caution and as illustrative. 

{\bf \cite{Nandra07}:} The first sample is taken from \cite{Nandra07} who examine 26 Seyfert-1 galaxies with public {\it XMM-Newton} data as of 2006 January 1, in order to characterize the Fe K$\alpha$ line. These authors restrict their sample to a minimum of 30,000 counts in the EPIC-pn between 2-10 keV for sufficient signal-to-noise to complete their study. After fitting the continuum and both narrow emission and absorption features, \cite{Nandra07} initially characterized the broad Fe line with a simple Gaussian. The authors then proceeded to use a more complex model for the broad features at 6.4 keV, a blurred reflection component, {\tt KDBLUR2 $\times$ PEXMON}. They fixed both the inner and outer radius of the accretion disk in this model, and let the break radius vary. For our study, we only utilize the broad Gaussian fits given in \cite{Nandra07}, for their complex model assumes either a non-rotating or maximally spinning black hole (i.e, no intermediate values of $a$). Of the 26 Seyferts, 15 met our detection standards and were used in our sample. See Table 2. The corresponding radio luminosity for each Seyfert is also listed in Table 2.

{\bf \cite{Patrick12}:} Our third sample is taken from \cite{Patrick12}, which uses {\it Suzaku} data which were public as of 2011 September. They require at least 30,000 counts in the 0.6--10 keV energy range of nearby Seyfert 1-1.9 galaxies (z$<$0.2). After the initial characterization of the continuum, which includes a power-law, a {\tt comptt} component modeling the soft excess, narrow emission lines, and warm-absorbers, the residuals in the Fe K band were characterized with a broad Gaussian feature. Of the 46 AGN in this sample, 19 of them have met our selection criterion. \cite{Patrick12} also characterized the excess in the Fe K band with a relativistic line model, {\tt RELLINE}. The relativistic line was frozen at 6.4 keV, and the emissivity, spin, inclination and flux of the line were all allowed to vary. We include 7 of the previous reported 19 Seyferts in our sample; these 7 have statistically significant detections of the broad Fe lines.  Finally, in addition to the radio luminosity for these sources, we also include the jet power proxy, $P_{Jet}$, described by \cite{Narayan12} as $P_{Jet} =D^2(\nu S_\nu)/M_{BH}$, where $D$ is the distance, $\nu$ is the observing frequency (5 GHz), $S_\nu$ is the flux density, and $M_{BH}$ is the mass of the black hole. See Table 2.

{\bf \cite{Walton12}:} The fourth sample is taken from \cite{Walton12}, who preform a uniform analysis of all {\it Suzaku} observations of Seyfert-1's available as of 2010 October. The authors choose their sample from Seyferts that have a soft excess and are ``bare" galaxies, which show little to no intrinsic absorption . The spectra are each characterized by a power-law component and a reflection component. The reflection component was modeled with {\tt REFLIONX}, a self-consistent reflection code, in conjunction with {\tt RELCONV}, a relativistic convolution kernel. These two codes characterize both the broad relativistic Fe K$\alpha$ line near 6.4 keV as well as the soft excess at $~$2 keV, thought to arise from the gravitationally blurred reflection off of the accretion disk. The Fe abundance, ionization, emissivity, inclination, spin and normalization were all allowed to vary in their analysis. A total of 16 Seyferts out of their sample of 25 were included in our work. Again we include the radio luminosity as well as the jet power. See Table 2.

{\bf Additional Seyferts:} We include an additional 6 Seyferts that have been studied individually. These are relatively nearby AGN ($z<0.07$), each of which have spin measurements from relativistic line modeling of the Fe K$\alpha$ line as well as compact radio emission. See Table \ref{tab:quant}. These Seyferts generally have high spin. However, 3C 120 shows evidence for low spin, $a<-0.1$ \citep{Cowperthwaite12}. \cite{Cowperthwaite12} do note that the X-ray observation was taken during a radio outburst and X-ray minimum, which may suggest the inner accretion disk had evacuated and thus the spin would be an lower limit. 

{\bf BHB:} Our final black hole sample is a compilation of spin and radio measurements of stellar-mass black holes. This sample includes those sources with spin measurements based on reflection and continuum modeling from a variety of different missions, including {\it XMM-Newton}, {\it Suzaku}, {\it RXTE}, {\it ASCA}, and {\it BeppoSAX}. Although other BHB spin measurements have been made using continuum fitting, we restrict the sample to just reflection or reflection in conjunction with continuum analysis for a consistent comparison to the the AGN sample. In addition, compact radio emission in BHB's can vary on week-to-month timescales. Because of this variability, we only include the highest radio flux listed in the literature as the peak luminosity of these sources. However, these may not necessarily be the peak luminosity needed to accurately compare to their AGN counterpoints, as not all of these sources have been sufficiently sampled in time. In addition, when the peak luminosity is measured at a wavelength other than 5 GHz, we used the documented spectral index or $\alpha=-0.3$ when unavailable, which is consistent with the median of our sample of BHB. Finally, the distance and mass to several of the sources is unknown, therefore we assumed a distance of 8 kpc and 10 $M_\odot$. See Table 2.

{\bf Neutron Stars:} Finally, we include a survey of neutron stars as our last sample, as a means of observationally probing lower spin values. Neutron stars cannot tap spin via magnetic field lines threaded through the ergosphere like black holes \citep{Blandford77}. However, in principle, they should be able to extract energy by spinning down the neutron star, analogous to the what may be happening in black holes. The sample itself is compiled directly from the work by \cite{Migliari06} and \cite{Migliari11}, which include X-ray fluxes, radio flux densities, distances and spin to neutron stars in several different X-ray states (Atoll, Z-sources, and millisecond X-ray pulsars). When more than one radio detection was available for a source, we took the highest radio flux density, which is consistent with our approach for the BHB. However, unlike the black hole spin measurements, the neutron star spin is based on the coherent millisecond pulsations, burst oscillations and QPO resonances. 

\subsection{Analysis}
We used three different statistical tests to examine the degree of correlation in each of our samples. We utilized the {\tt ASURV} code to implement both a Spearman's rank correlation test as well as the Kendall rank correlation test, which has the ability to account for both upper and lower limits in the samples \citep{Isobe86}. We used a code based on the methodology of \cite{Akritas96} to implement a partial correlation test, which could also handle censored data in our samples. 

The Spearman's rank correlation test  ranks the values of each variable in the sample and uses the difference between the ranks of the dependent and independent variables to estimate the correlation in the sample. The coefficient, $\rho_S$, which is used to quantify the correlation, spans the ranges $-1<\rho_S<1$ (where -1 is anti-correlated, 1 is positively correlated, and 0 denotes no correlation between the two variables). This test is most sensitive to monotonic distributions, and although it does not take uncertainties into account, our code was able to weight censored data less than detected data.

The Kendall rank correlation test, which is also a rank correlation test like the Spearman's rank correlation test, is a ranking test which is based on whether a pair of variables ($x_i,y_i$) is correlated with other pairs of variables, (e.g., $x_i>x_j,y_i>y_j$). It is a relative ranking scheme rather than an absolute ranking of the the variables in the Spearman's rank correlation test. The Kendall's $\tau_K$ coefficient also describes how prominent the correlation is in the sample, just as the Spearman's rank correlation test does, where $\tau_K$ spans $-1<\tau_K<1$ and -1 implies anti-correlation, 0 implies no correlation, and 1 implies positive correlation. In general, these test agree with each other, although the magnitude of $\tau_K$ and $\rho_S$ cannot strictly be compared with one another due to the different algorithms. In addition, in small samples (n$<$20), such as presented in this study, the Kendall's $\tau_K$ is preferred over the Spearman's $\rho_S$ because of use of relative rankings between pairs when calculating $\tau_K$ versus the absolute ranking of the small sample to calculate the $\rho_S$.

Our final statistical test, the partial correlation test, is a variant of the Kendall rank correlation test which quantifies the correlation in a data set given a third variable. This is done utilizing a particular combination of the Kendall's $\tau_K$ calculated between each of the three variables being considered. Further, by implementing the work by \cite{Akritas96} that describes a censored partial correlation test, we are not only able to use the partial correlation test on detections, but also those measurements that have upper or lower limits. By utilizing a partial correlation test, we are able to understand whether additional variables have a strong influence on the observed correlations. This is most important when trying to understand the influence of distance on these relations, which can influence the flux and therefore net counts and signal-to-noise in the X-ray spectra. All confidence levels quoted are for the partial correlation test, unless otherwise stated. 

	\begin{figure*}[t]
	\centering
	\subfigure[This plot shows the  Fe K$\alpha$ Gaussian line width versus the Doppler-corrected radio luminosity for two different samples of Seyfert galaxies. \cite{Patrick12} sample is in black, and \cite{Nandra07} sample is in red. No statistically significant correlation is observed. \label{fig:sigvr}]{
		\includegraphics[scale=.4,angle=0,clip=true,trim=1 1 1 1]{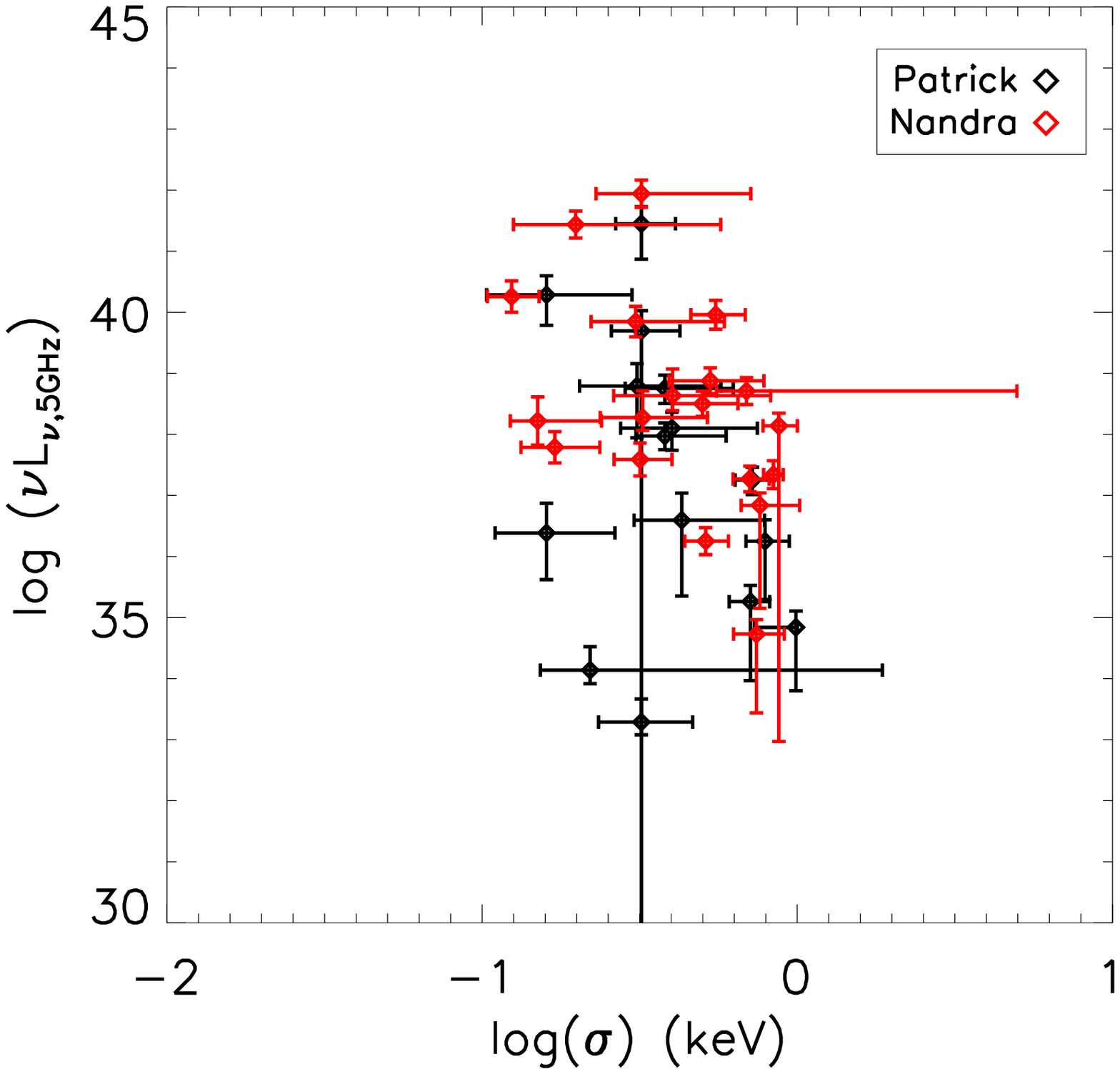}
	}\hspace{1cm}
	\subfigure[This plot displays the Gaussian line widths vs the Doppler-corrected radio luminosity per unit mass in natural units. The color scheme is the same as Figure \ref{fig:sigvr}. We note there is no statistically significant correlation present. \label{fig:sigvpjet}]{
		\includegraphics[scale=.4,angle=0,clip=true,trim=1 1 1 1]{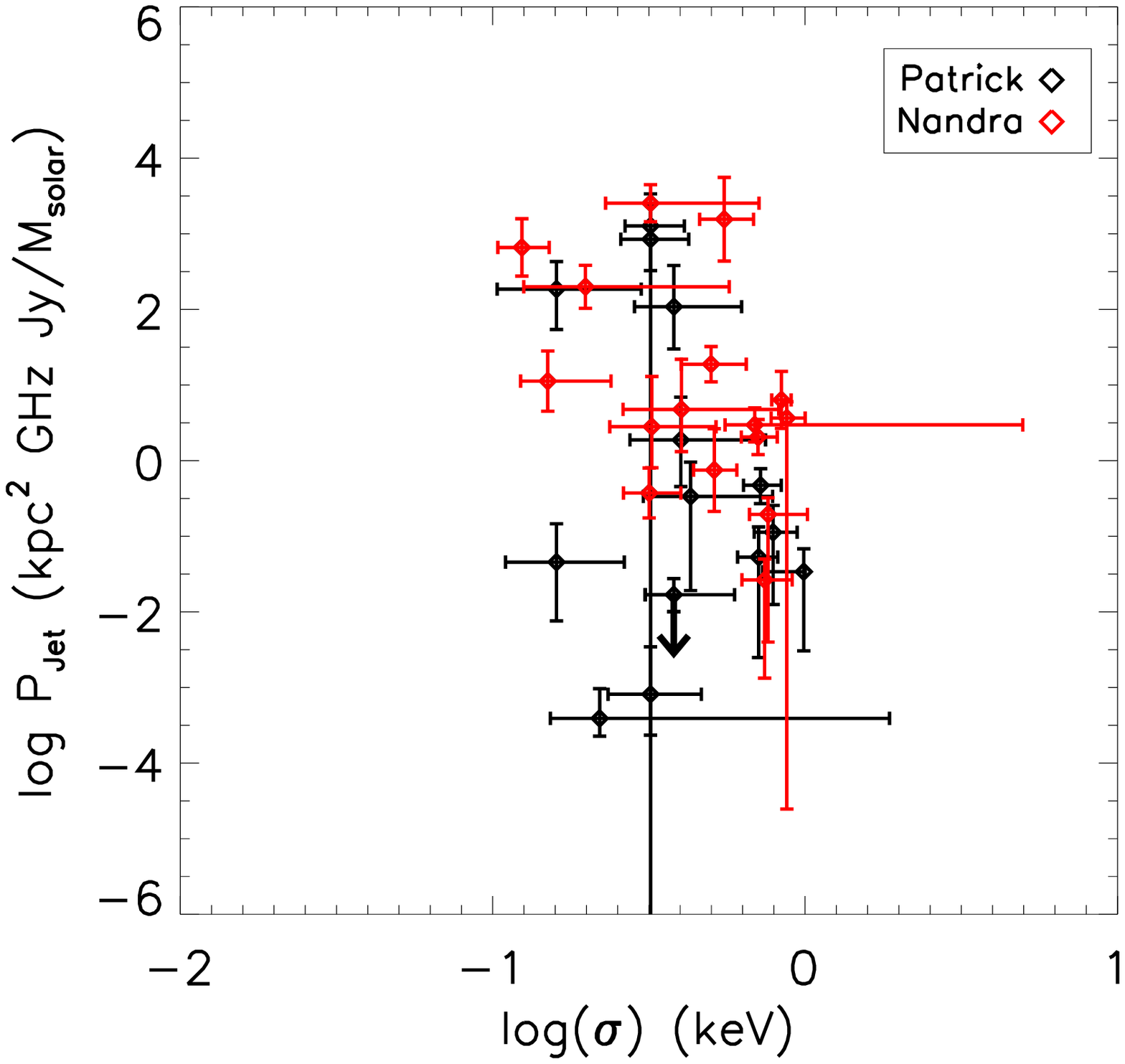}
	}
\subfigure[This plot shows the Gaussian line widths vs the Doppler-corrected radio luminosity per unit mass squared. This scaling is suggested by \cite{Blandford77}. The color scheme is the same as Figure \ref{fig:sigvr}. There is no statistically significant correlation present. \label{fig:sigvm2}]{
\includegraphics[scale=.4,angle=0,clip=true,trim=1 1 1 1]{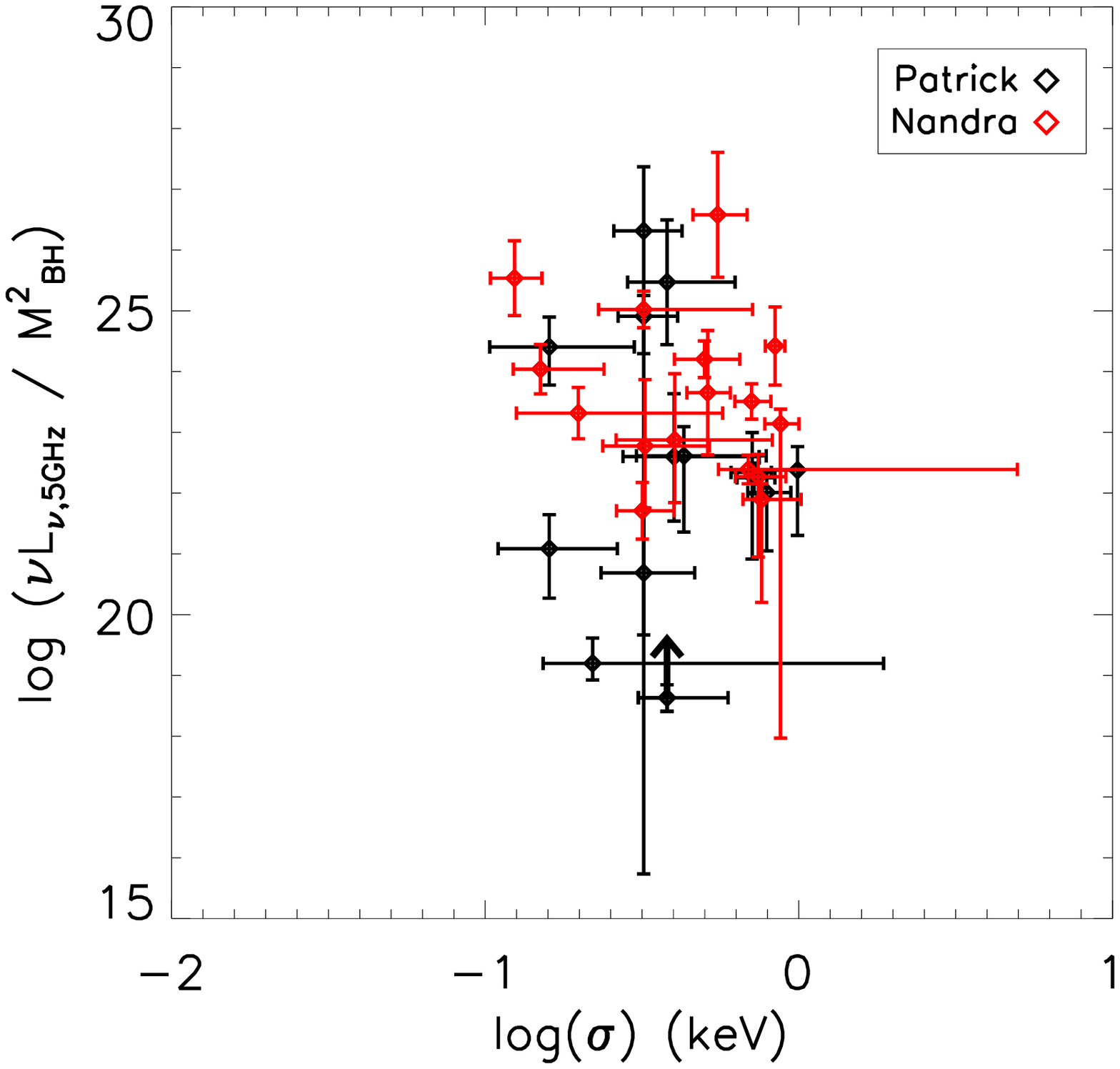}
}\hspace{1cm}
\subfigure[The above plot shows the Gaussian line widths versus the Doppler-corrected radio luminosity per Bolometric luminosity, as a means of assessing the dependence of mass accretion rate in this sample. There is no statistically significant correlation. \label{fig:sigvrb}]{
		\includegraphics[scale=.4,angle=0,clip=true,trim=1 1 1 1]{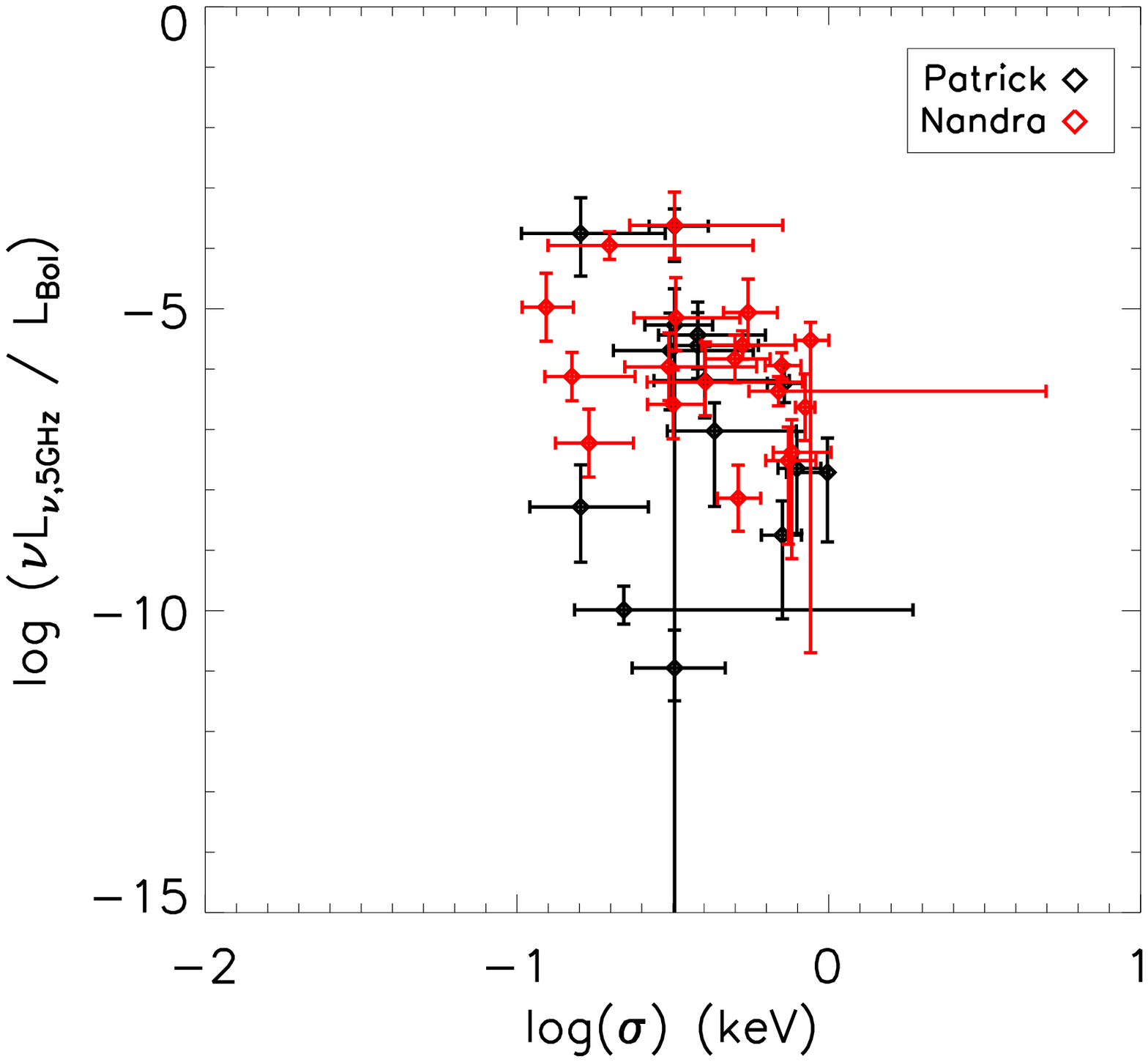}
	}

	\caption{}
	\end{figure*}

\begin{deluxetable*}{l l l |l l l | l l | l l }[h]
\tablecolumns{10}
\tablewidth{0pc}
\tabletypesize{\scriptsize}
\tablecaption{Correlation Tests}
\tablehead{ Sample & Size & correlated & Partial & Corr & & \multicolumn{2}{l}{Kendall's}&\multicolumn{2}{l}{Spearman's} \\
&&& $\tau_{p}$ &$\sigma_p$ & Prob & $\tau_K$ & Prob & $\rho_s$ & Prob  }
\startdata
\cite{Nandra07} &  15 &$\log \sigma$ vs $\log \nu L_R$ (z) &  -0.176 & 0.229 & 0.442 & -0.30 & 0.111  & -0.32 & 0.238 \\
\cite{Patrick12} &  19 &$\log \sigma$ vs $\log \nu L_R$ (z)  & -0.119 & 0.103 & 0.248& -0.31 & 0.064 & -0.49 & 0.036\\
ALL$^1$ &  26 & $\log \sigma$ vs $\log \nu L_R$ (z) &- 0.189 & 0.137 &0.168 & -0.26 & 0.048 & -0.35 & 0.066  \\
\\
\cite{Nandra07} & 14 & $\log \sigma$ vs $\log P_{Jet}$ (z) & -0.047 & 0.204 & 0.818 & -0.15 & 0.389 & -0.12 & 0.656 \\
\cite{Patrick12} &  16 & $\log \sigma$ vs $\log P_{Jet}$ (z) & -0.149 & 0.126 & 0.237 & -0.30 & 0.105 & -0.45 & 0.083\\
ALL & 21 & $\log \sigma$ vs $\log P_{Jet}$ (z) & -0.187 & 0.144& 0.194 & -0.29 & 0.036 & -0.40 & 0.061\\
\\
\cite{Nandra07} &  14 &$\log \sigma$ vs $\log  (\nu L_R/M^2_{BH})$ (z) & -0.017 & 0.173 & 0.921 & -0.07 & 0.712 & -0.01 & 0.971\\
\cite{Patrick12} &  16 &$\log \sigma$ vs $\log  (\nu L_R/M^2_{BH})$ (z)  & -0.112 & 0.185 & 0.545 & -0.18 & 0.322 & -0.23  & 0.380\\
ALL$^1$ &  21 & $\log \sigma$ vs $\log (\nu L_R/M^2_{BH})$ (z) & -0.183 & 0.157 & 0.244 & -0.26 & 0.062 & -0.35 & 0.104\\
\\
\cite{Nandra07} & 15 & $\log L_X$ v $\log \nu L_R$ (z) &0.130 & 0.146 & 0.373& 0.30 & 0.126 & 0.41 & 0.129 \\
\cite{Patrick12} & 19 &$\log L_X$ v $\log \nu L_R$ (z)&0.226 & 0.194 & 0.244 & 0.6  & 0.00032 & 0.80 & 0.00065 \\
ALL & 27 & $\log L_X$ v $\log \nu L_R$ (z) &0.156 & 0.113 & 0.167 & 0.48 & 0.00064  & 0.66 & 0.00091\\
\\
\cite{Nandra07} &  15 &$\log \sigma$ vs $\log  (\nu L_R/L_{Bol})$ (z) &  -0.135 & 0.226 & 0.55 & -0.23 & 0.232 & -0.19 & 0.489\\
\cite{Patrick12} &  19 &$\log \sigma$ vs $\log  (\nu L_R/L_{Bol})$ (z)  & -0.088 & 0.136 & 0.518& -0.19 & 0.248 & -0.30 & 0.198\\
ALL$^1$ &  26 & $\log \sigma$ vs $\log  (\nu L_R/L_{Bol})$ (z) &-0.185 & 0.160 & 0.248 & -0.22 & 0.064  & -0.29 & 0.098\\
\\

Seyfert Sample& 21 &  $\log |a|$ vs $\log \nu L_R$ (z) &0.034 & 0.085 & 0.689 & 0.05 & 0.618 &  0.16 & 0.489 \\
BHB Sample& 11 &  $\log |a|$ vs $\log \nu L_R$ (z) &0.454 & 0.196 & 0.021  & 0.47 & 0.042 & 0.65 & 0.040\\
Neutron Stars & 13 & $\log |a|$ vs $\log \nu L_R$ (d) &-0.409 & 0.170 & 0.016 & -0.46 & 0.027&0.-0.65 & 0.024 \\

\\

Seyfert Sample & 17 &  $\log |a|$ vs $\log P_{Jet}$ (z) &  0.123 & 0.089 & 0.167 & 0.15 & 0.219 & 0.35 & 0.162 \\
BHB Sample & 11 &  $\log |a|$ vs $\log P_{Jet}$ (z) &  0.418 & 0.208 & 0.044& 0.43 & 0.012 & 0.063 & 0.061\\
Black Holes & 28 &$\log |a|$ vs $\log P_{Jet}$ (z) & 0.225 & 0.089 & 0.011 & 0.28 & 0.010 & 0.49 & 0.012\\
Neutron Stars & 13 & $\log |a|$ vs $\log P_{Jet}$ (d) &-0.426 & 0.163 & 0.0089 & -0.46 & 0.024 & -0.68 & 0.019\\
{\bf ALL  Sources} &41 & {\bf $\log |a|$ vs $\log P_{Jet}$ (d)} & {\bf 0.219 } & {\bf 0.066}& {\bf 0.00091 }& {\bf 0.35} &{\bf  0.00025} &{\bf  0.60} & {\bf 0.0010}\\
\\
Seyfert Sample & 17 &$\log |a|$ vs $\log  (\nu L_R/M_{BH}^2) (z)$ & 0.124 & 0.090 & 0.168 & 0.11 & 0.350 & 0.26 & 0.298 \\
BHB Sample & 11 &$\log |a|$ vs $\log  (\nu L_R/M_{BH}^2) (z)$ & 0.418 & 0.208 & 0.044 & 0.44  & 0.061 & 0.63 & 0.047\\
Black Holes &28 &$\log |a|$ vs $\log  (\nu L_R/M_{BH}^2) (z)$ & 0.107 & 0.070 & 0.126  & 0.02 & 0.868 & 0.07 & 0.710\\
Neutron Stars &13& $\log |a|$ vs $\log (\nu L_R/M_{BH}^2)$ (d) &-0.426 & 0.163 & 0.0089& -0.46 & 0.024 & -0.68 & 0.019 \\
ALL Sources & 41 & $\log |a|$ vs $\log (\nu L_R/M_{BH}^2)$ (d) &-0.173 & 0.101 & 0.087 & -0.28 & 0.0041& 0.-40 & 0.010 \\
\\
Seyfert Sample & 21 & $\log |a|$ vs $\log  (\nu L_R/L_{Bol})$ (z) & 0.053 & 0.086 & 0.537  & 0.06 & 0.553 & 0.16 & 0.479 \\
BHB Sample & 9 & $\log |a|$ vs $\log  (\nu L_R/L_{Bol})$ (z) &  0.479 & 0.203 & 0.018& 0.19 & 0.289 & 0.31 & 0.262  \\
Black Holes & 30 & $\log |a|$ vs $\log  (\nu L_R/L_{Bol})$ (z) & 0.164 & 0.091 & 0.072  & 0.19 & 0.057 & 0.34 & 0.070  \\
Neutron Stars & 13 &  $\log |a|$ vs $\log (\nu L_R/L_{Bol})$ (d)& 0.402& 0.154& 0.009& 0.42 & 0.035 & 0.58 & 0.045 \\
{\bf ALL Sources}&  45 &  {\bf $\log |a|$ vs $\log (\nu L_R/L_{Bol})$ (d)}& {\bf 0.391} &{\bf  0.089} & {\bf 1.1$\times10^{-5}$ }& {\bf 0.49 }&{\bf 1.0 $\times10^{-7 }$} & {\bf 0.75} & {\bf 7.5 $\times10^{-7}$}\\
\\
Seyfert Sample & 17 & $\log |a|$ vs $\log \frac{\nu L_R}{L_{Bol}^{\alpha}M_{BH}^{\beta}}$ (z) &  0.088 & 0.108 & 0.415 & 0.07 & 0.540 & 0.20 & 0.419\\
BHB Sample & 9 & $\log |a|$ vs $\log   \frac{\nu L_R}{L_{Bol}^{\alpha}M_{BH}^{\beta}}$ (z) &   0.398 & 0.222 & 0.073 & 0.42 & 0.116 & 0.63 & 0.076\\
Black Holes & 26 & $\log |a|$ vs $\log  \frac{\nu L_R}{L_{Bol}^{\alpha}M_{BH}^{\beta}}$ (z) & 0.189 & 0.104 & 0.069 & 0.17 & 0.131 & 0.32  & 0.114 \\
Neutron Stars & 13 &  $\log |a|$ vs $\log \frac{\nu L_R}{L_{Bol}^{\alpha}M_{BH}^{\beta}}$ (d)&  0.118 & 0.184 & 0.521 & 0.10 & 0.589 & 0.15 & 0.609\\
{\bf ALL Sources}&  39 &  {\bf $\log |a|$ vs $\log  \frac{\nu L_R}{L_{Bol}^{\alpha}M_{BH}^{\beta}}$ (d)}& {\bf 0.344} & {\bf 0.083} & {\bf 3.4$\times10^{-5}$} & 0.23 & 0.018 & 0.37 & 0.023\\
\enddata  
\label{tab:corr}
\tablecomments{\small{This table shows the correlation tests of line width ($\sigma$), spin ($a$), and X-ray luminosity ($L_X$) versus numerous jet power proxies \citep[$\nu L_R$, $P_{Jet}$, $\nu L_R/M^{2}_{BH}$, $\nu L_R/L_{Bol}$, $\nu L_R/L_{Bol}^{\alpha}/M^\beta_{BH}$, where $\alpha = 0.78$ and $\beta = 0.67$, see][]{Gultekin09}. The radio luminosity are all Doppler-corrected except for the Neutron stars. There is no statistically significant trend when using the Guassian line widths, although a tentative inverse correlation is observed. However, the correlations in bold are statistically significant ($>3\sigma$), positive trends when comparing spin measurements to $P_{Jet}$, $\nu L_R/L_{Bol}$, and $\nu L_R/L_{Bol}^{\alpha}/M^\beta_{BH}$, and employing the entire compact object samples.} }

\end{deluxetable*} 
\begin{figure}[h]
\centering
\hspace{1cm}
	\subfigure[ This plots shows the Gaussian line width as a function of redshift. \cite{Patrick12} sample is in black, and the \cite{Nandra07} sample is in red. There is no apparent dependence on redshift. \label{fig:zvsig} ]{
		\includegraphics[scale=.4,angle=0,clip=true,trim=1 1 1 1]{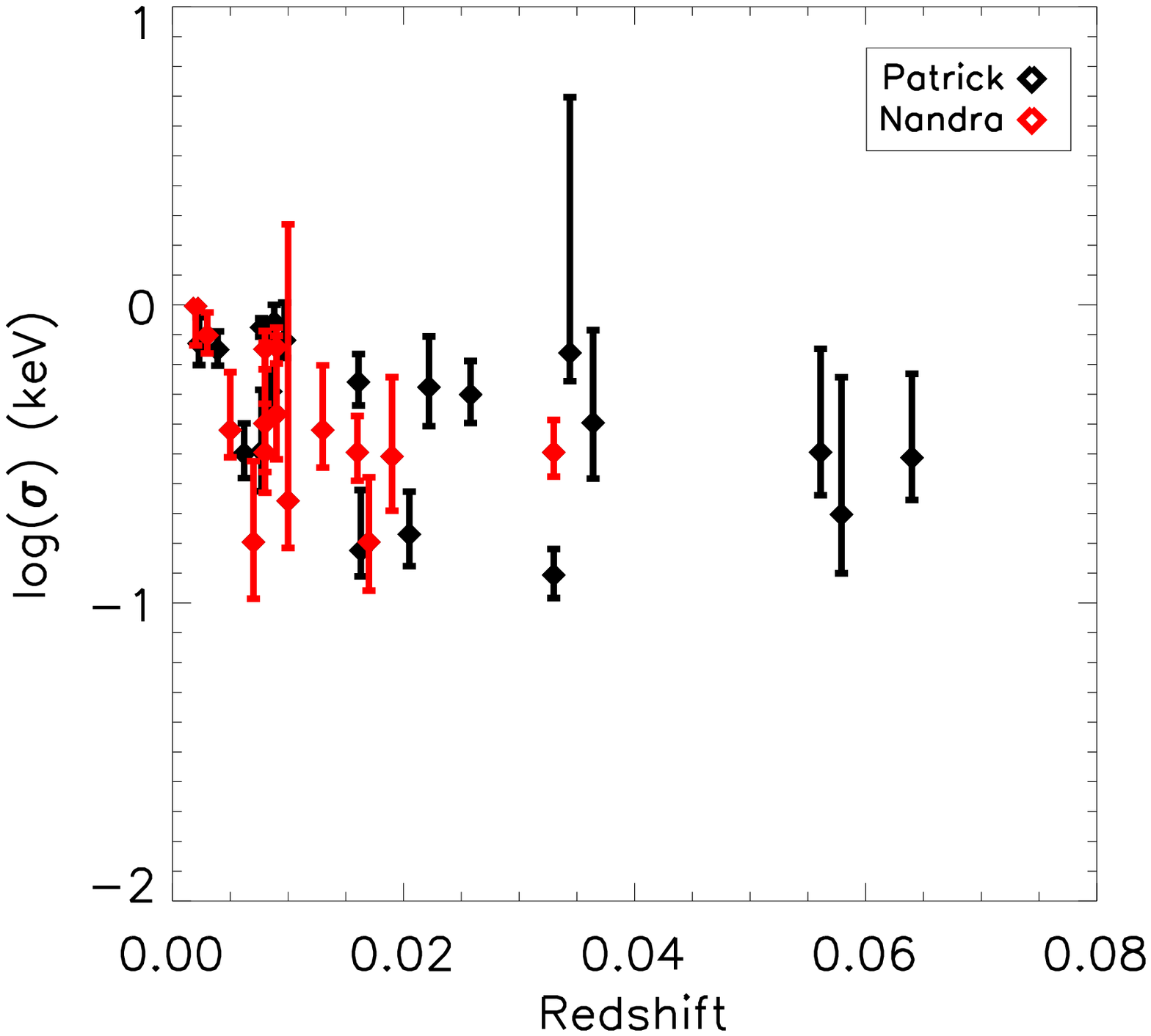}
	}\hspace{1cm}

\hspace{1cm}
	\subfigure[The above figure plots the absolute value of spin as a function of redshift. The \cite{Walton12} sample is in black, and our 5 additional Seyferts are in cyan. There does not appear to be a spin dependence on redshift. \label{fig:zva} ]{
		\includegraphics[scale=.4,angle=0,clip=true,trim=1 1 1 1]{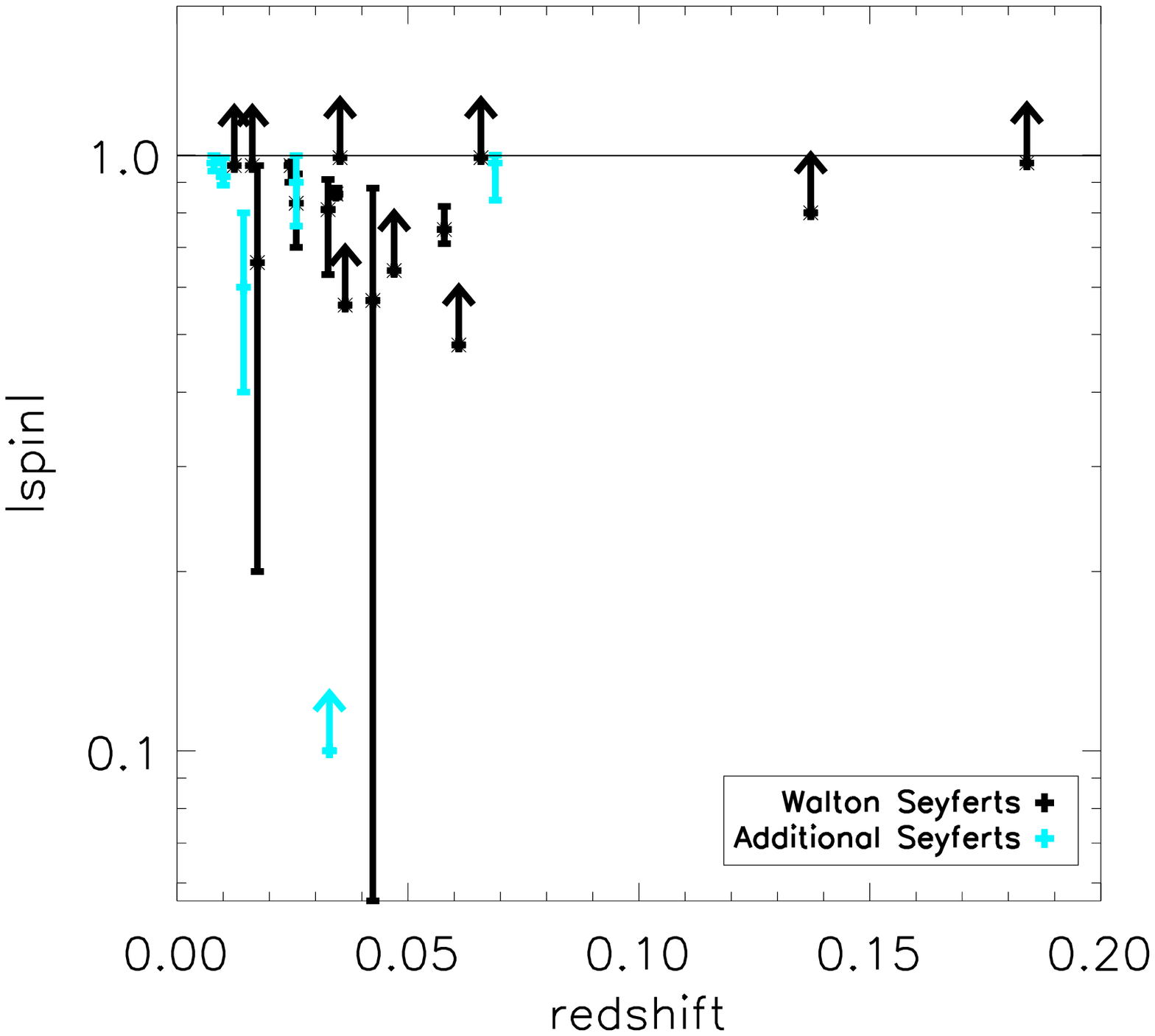}
	}
\caption{}
\end{figure}

\begin{figure}[h]
\centering
\includegraphics[scale=.4,angle=0,clip=true,trim=1 1 1 1]{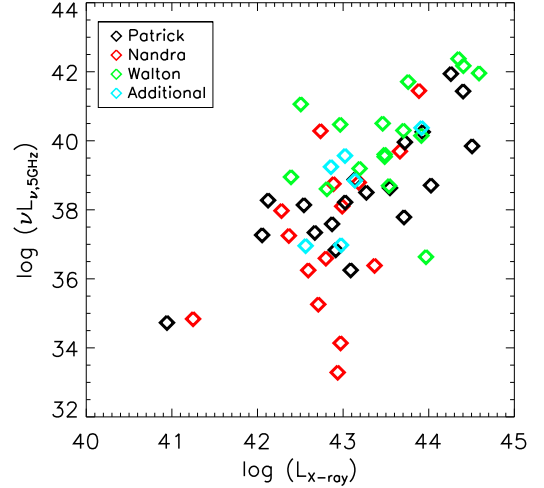}
\caption{This plot shows the Doppler-corrected radio luminosity vs. X-ray luminosity of all our Seyfert samples.  \cite{Patrick12} is in black, \cite{Nandra07} is in red, \cite{Walton12} is in green and our 5 additional Seyferts are in cyan. These sources show a positive correlation between the mass accretion rate (X-ray) and jet power (radio). \label{fig:xvr}}
\end{figure}

\begin{figure*}[t]
\centering
	\subfigure[This plot shows the absolute value of the Doppler-corrected radio luminosity versus the spin of our composite Seyfert Sample. The points in cyan have inclinations that were frozen.  There is a tentative positive correlation. The dashed line is $\nu L_\nu \propto a^2$ \label{fig:avr}]{	
		\includegraphics[scale=.36,angle=0,clip=true,trim=1 1 1 1]{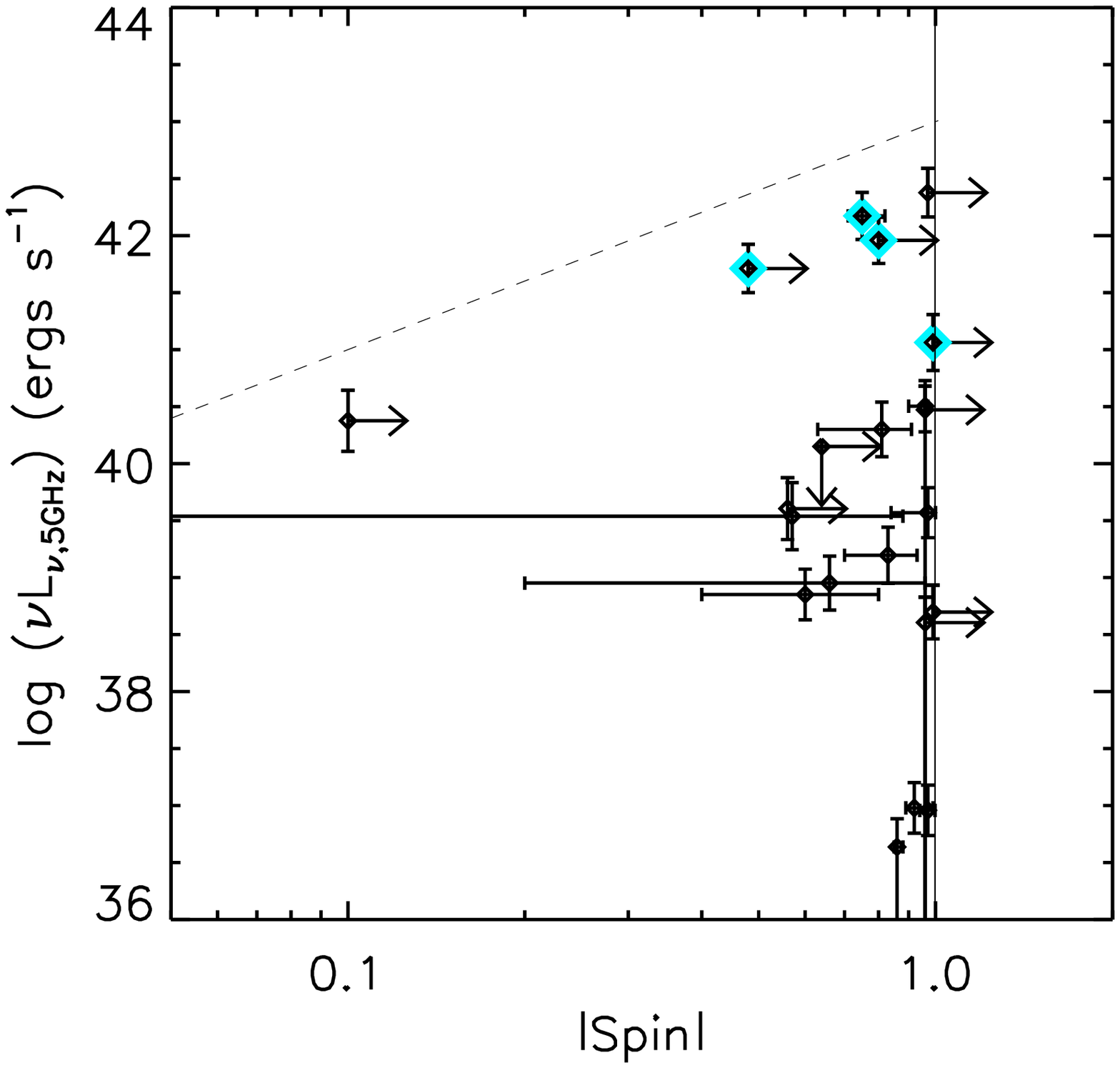}
	}\hspace{1cm}
	\subfigure[This figure compares the absolute value of the Doppler-corrected radio luminosity per unit mass in natural units to the spin parameter. The dashed line is $P_J \propto a^2$. The black points are Seyferts, the blue points are BHB and the red points are neutron stars. When considering the entire compact object sample, we find a partial correlation coefficient of $\tau_p=0.219$ with a 3.3$\sigma$ confidence level of correlation. \label{fig:avpjet}]{
\includegraphics[scale=.36,angle=0,clip=true,trim=1 1 1 1]{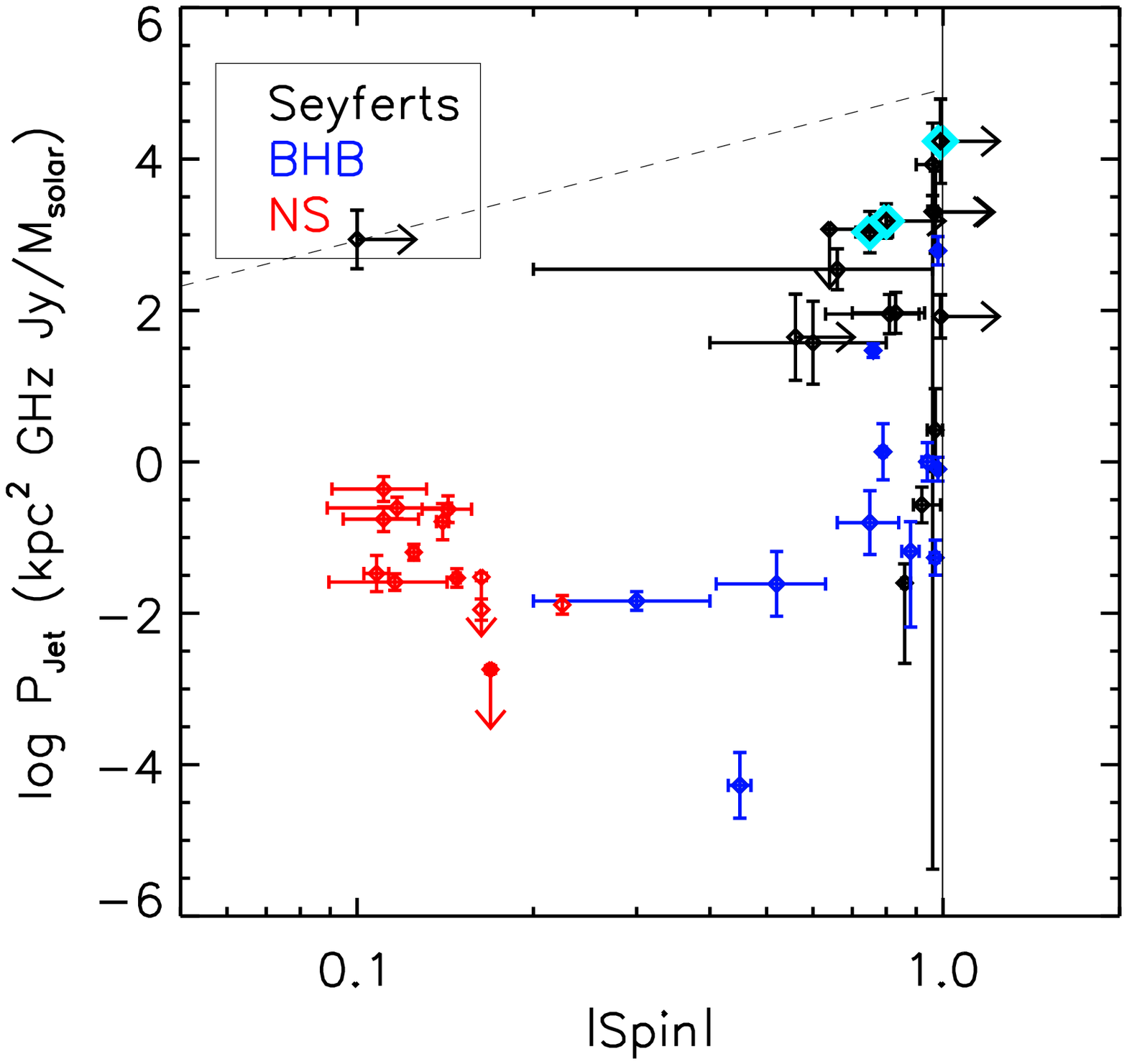}
	}
	\subfigure[This plots the absolute value of the Doppler-corrected radio luminosity per unit mass$^2$ versus the spin parameter. The dashed line shows $\nu L_{\nu,5GHz}/M^2_{BH} \propto a^2$ . The sources are stratified by mass with the BHB and neutron stars being approximately 6 orders of magnitude higher than the Seyferts. This can roughly be explained by equipartition of magnetic field strength in these sources if, $B^2 \propto L_{Bol}/R^2$. \label{fig:avm2}]{
\includegraphics[scale=.36,angle=0,clip=true,trim=1 1 1 1]{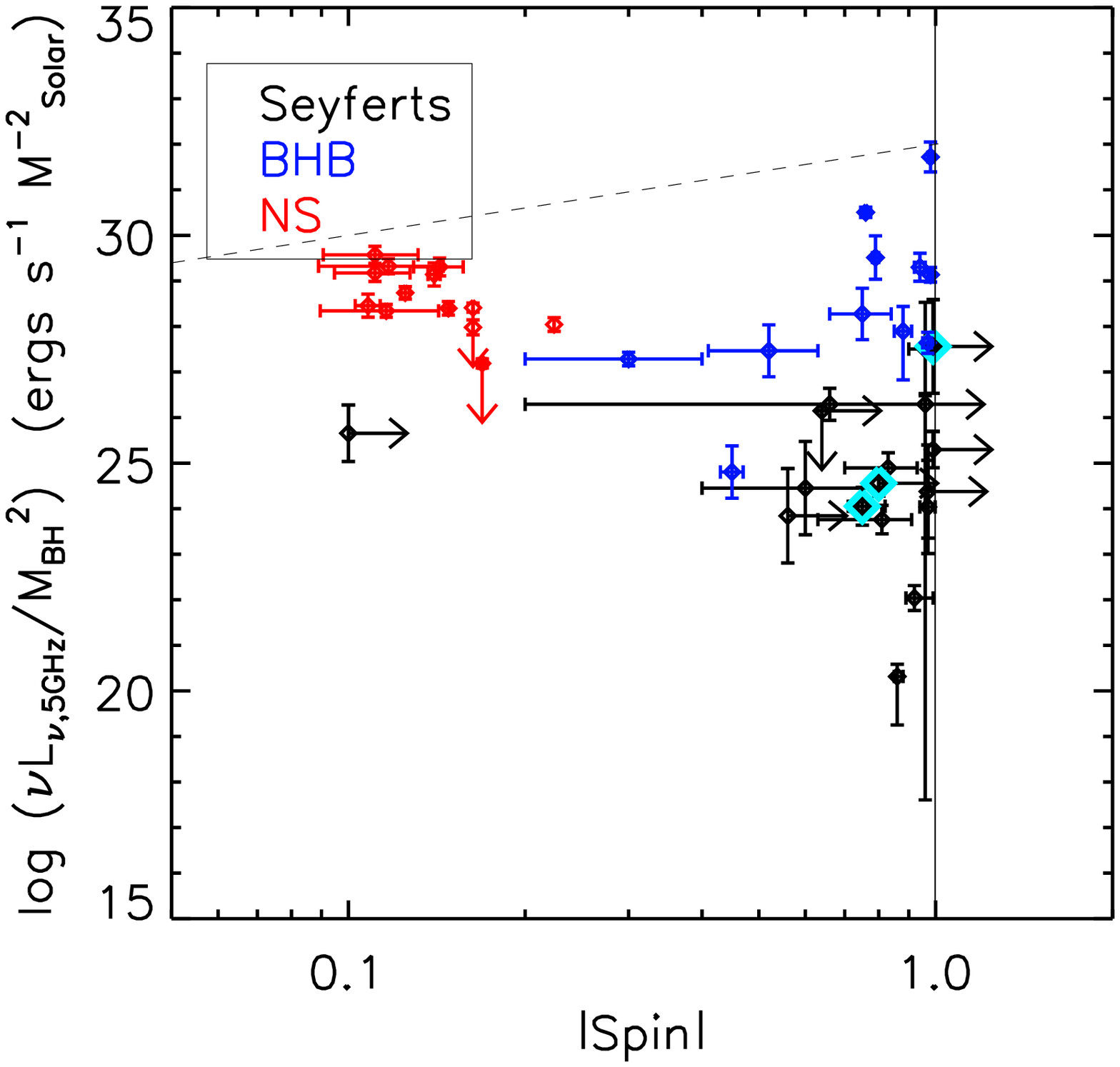}
}\hspace{1cm}
\subfigure[This plots shows the absolute value of the Doppler-corrected radio luminosity per Bolometric luminosity versus the spin parameter. The data is positively correlated with $\tau_p$=0.391 at a 4.4$\sigma$ confidence level. This suggests that spin has a dramatic role in determining the jet luminosity as does mass accretion rate (i.e. Bolometric luminosity) . The scatter is likely driven by the non-simultaneity and Bolometric correction of the X-ray data in the Seyfert sample. \label{fig:avrb}]{
\includegraphics[scale=.36,angle=0,clip=true,trim=1 1 1 1]{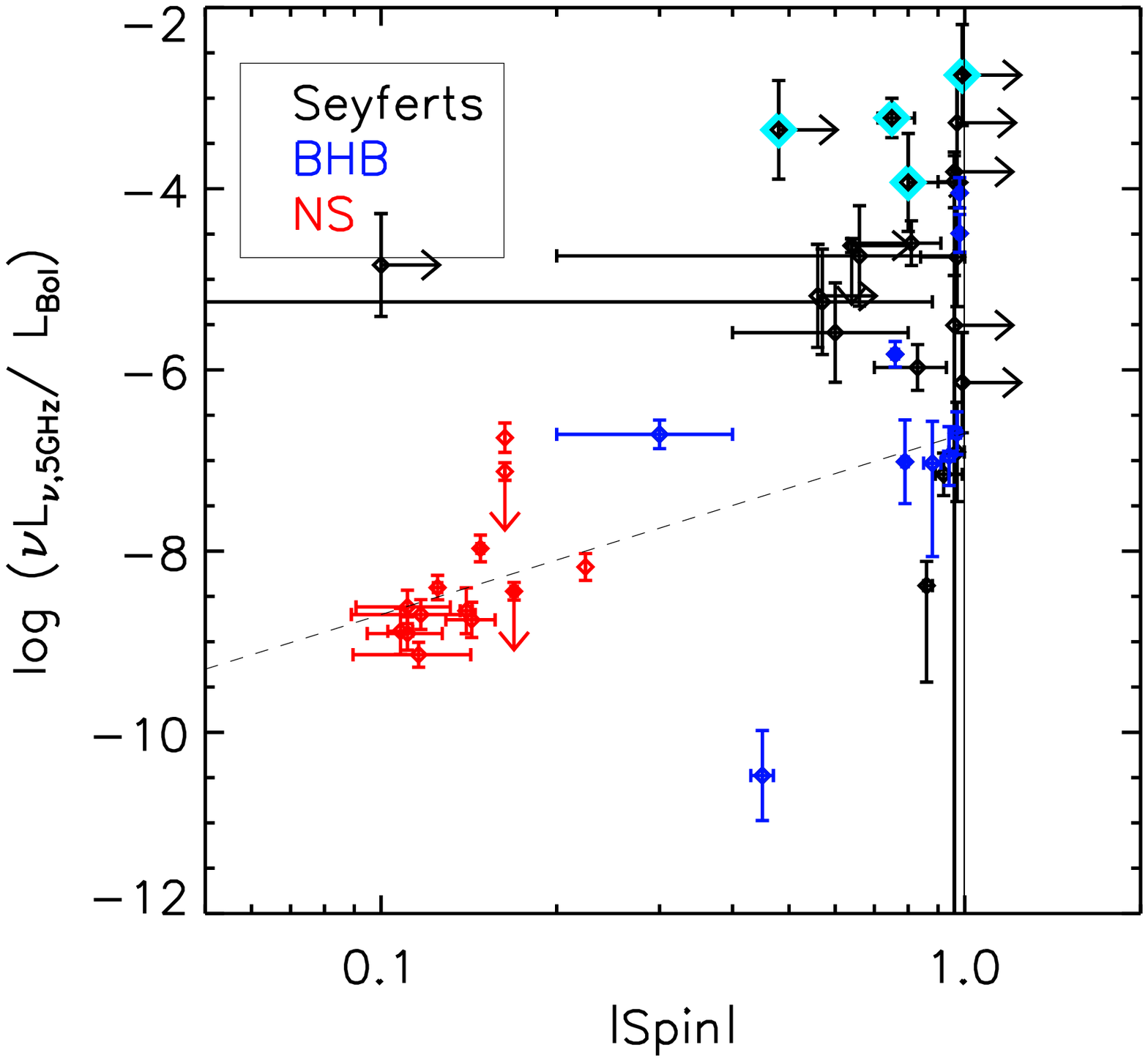}}
\end{figure*}

\begin{figure*}
\hspace{1cm}
\subfigure[This is the same plot as Figure \ref{fig:avrb} but only including BHB and neutron stars to demonstrate the tighter correlation in the stellar-mass sources as compared to the Seyferts. \label{fig:avrbbhb}]{
\includegraphics[scale=.36,angle=0,clip=true,trim=1 1 1 1]{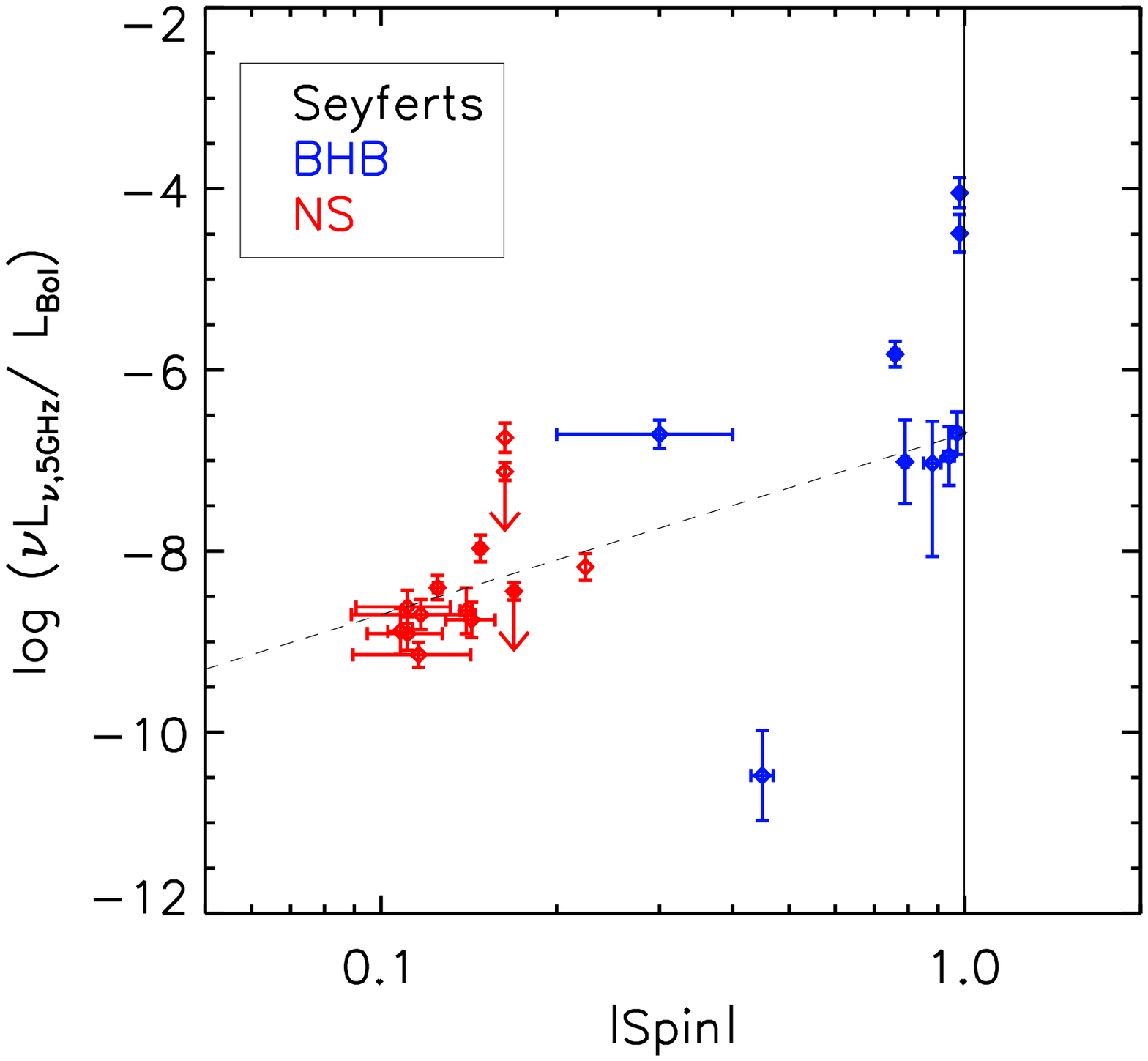}
}\hspace{1cm}
\subfigure[This figure shows the $\log  (\nu L_R/L_{Bol}^{0.67}/M_{BH}^{0.78})$ versus the absolute spin measurements . This is following the Fundamental plane of black hole accretion given by \cite{Gultekin09}. Even when correcting for mass and Bolometric luminosity there is still a statistically significant (4.1$\sigma$) correlation with spin (examining the compact samples as a whole).  \label{fig:avfund}]{
\includegraphics[scale=.36,angle=0,clip=true,trim=1 1 1 1]{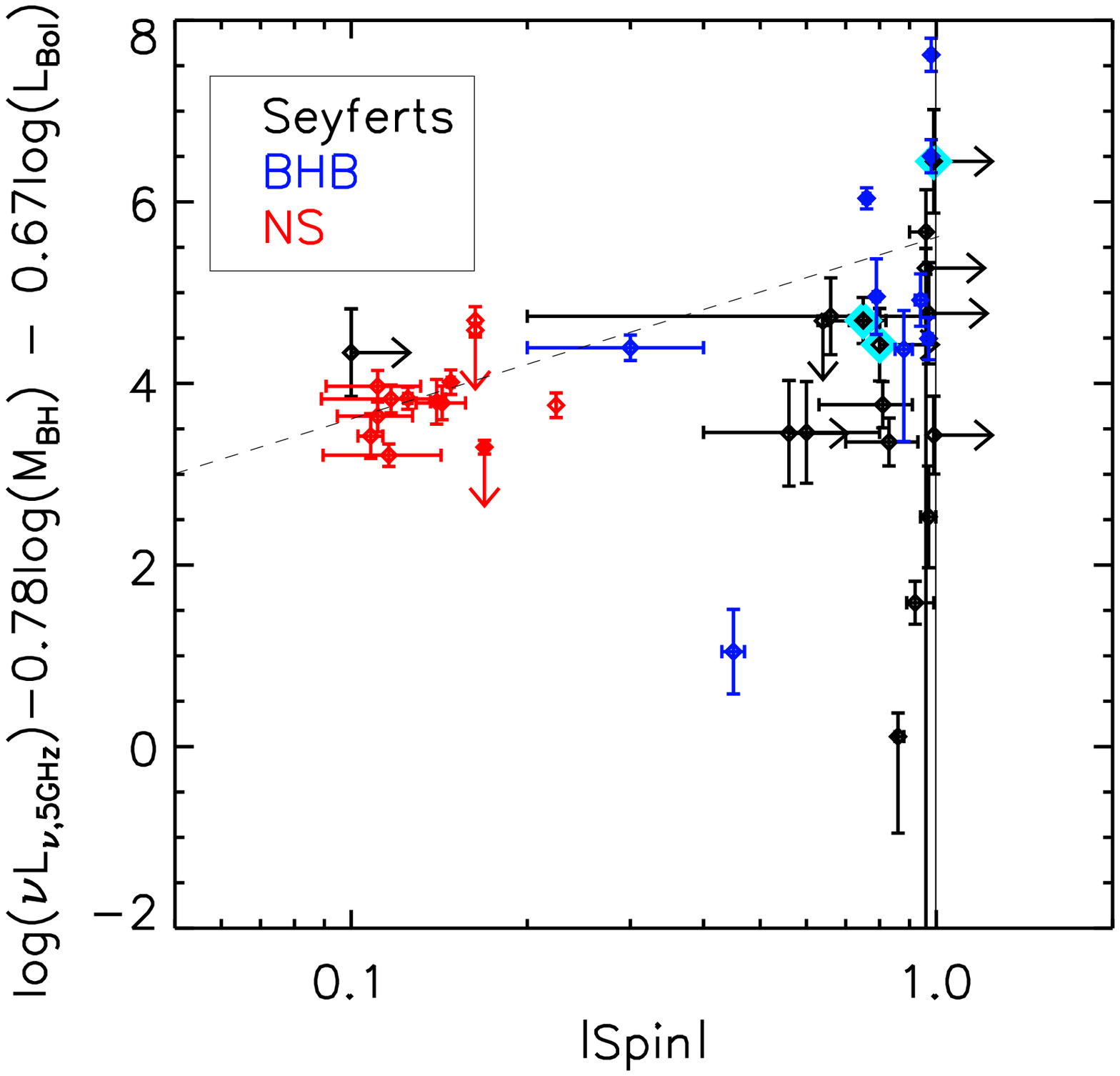}
}
\hspace{1cm}
\caption{}
\end{figure*}

\section{Results}
\label{disc}
\subsection{Gaussian Fits}
We have compiled two different Seyfert samples that 1) show excess in the Fe K$\alpha$ region, 2) are modeled with Gaussian line fits, and 3) have documented compact radio emission. The assumption in this analysis is that the Fe lines are produced in the inner accretion disk and emit at the inner-most stable circular orbit (ISCO). The ISCO is set by the spin of the black hole, 1.24 R$_G$ for maximally prograde holes, 6 R$_G$ for Schwarzschild holes, and 9 R$_G$ for a maximally spinning retrograde holes \citep{Bardeen72}. Consequently, the emission line shape broadens as the spin of the black hole increases, due to the material emitting from deeper in the potential well, i.e. closer to the black hole \citep{Miller07}. We also assume that the radio emission is a proxy for jet power. \cite{Merloni07} show a positive correlation between jet power as measured via X-ray cavities and radio luminosity, though with broad scatter. This trend allows us to use radio luminosity, which is much more readily available for each galaxy as compared to X-ray cavities, which may be too faint to observe in many AGN. 

Figures \ref{fig:sigvr}, \ref{fig:sigvpjet} \& \ref{fig:sigvm2} show the composite samples comparing the Fe K$\alpha$ line widths versus three different jet power proxies. The plots show all of the data listed in Table \ref{tab:quant}, but in the correlation tests that include both samples, each Seyfert is counted only once with preference to \cite{Nandra07} over \cite{Patrick12}. When comparing the Doppler-corrected radio luminosity to Gaussian line width, as in Figure \ref{fig:sigvr}, we are assuming that each of the Seyferts can potentially power the jets with the same magnitude. However, the range in Seyfert mass spans over 3 orders of magnitude, as show in Table \ref{tab:quant}. We might expect that the more massive the black hole, the more powerful the jet it can produce, based on the potential of the black hole, alone. Therefore, in Figure \ref{fig:sigvpjet} we plot the radio luminosity per unit black hole mass ($P_{Jet}$) versus line width. In particular, we use the relation given in \cite{Narayan12}, 
\begin{equation}
P_{Jet} = D^2 (\nu S_{\nu,5GHz}) / M_{BH} \ ({\rm kpc^2 \  GHz \ Jy \ M_\odot^{-1}})
\end{equation}
to estimate this proxy for the jet power, where $D$ is the distance to the source, $\nu$ is the observing frequency taken to be 5 GHz, $S_{\nu,5GHz}$ is the Doppler-corrected radio flux density at 5 GHz, and $M_{BH}$ is the mass of the black hole. 

In Figure \ref{fig:sigvm2}, we plot the Doppler-corrected radio luminosity per unit black hole mass squared versus line width. This is derived from the jet power estimate given by \cite{Blandford77}, who predict the jet luminosity should scale as $L_{Jet}\propto a^2B^2 M_{BH}^2$. Unfortunately, we do not have an estimate for the magnetic field in each source, so this could be an extra source of scatter in the relation.

Each of these three jet power proxies displays a slight inverse correlation between the line widths and radio luminosity per $M_{BH}^{0}$, $M_{BH}^{1}$, \& $M_{BH}^{2}$. This is also observed in the statistical tests we applied to each scenario (See Table \ref{tab:corr}). The redshift dependence of the line widths is also plotted in Figure \ref{fig:zvsig}.

We also plot the  Doppler-corrected radio luminosity per Bolometric luminosity versus line width as a way of assessing the dependence on mass accretion rate in Figure \ref{fig:sigvrb}. We have used the X-ray luminosity as well as a Bolometric correction as a proxy for mass accretion rate. The Bolometric correction factor is either taken directly from \cite{Vasudevan09} or assumed to be $\kappa = 20$ derived from \cite{Vasudevan09}. Again, we observe only a tentative negative correlation.

For comparison, we also include the correlation between the  Doppler-corrected radio luminosity and X-ray luminosity shown in Figure \ref{fig:xvr}. If the X-rays are assumed to be a proxy for the mass-accretion rate, this plot represents the correlation between the mass-accretion rate and the jet power in these systems. As shown in Table \ref{tab:corr}, there is statistically significant positive correlation when employing either the Kendall's $\tau$ or Spearman's rank correlation test, but only a tentative positive correlation when we take redshift into account. However, previous studies of this particular relation with larger samples show statistically significant correlation that extend not just to Seyferts but to other AGN and stellar-mass black holes as well \citep[e.g.,][]{Merloni03,Falcke04,Gultekin09}. We refer the reader to these studies for further insight on this particular correlation.

\subsection{Spin Measurements from Compact Objects}
\label{res2}
Although there are many Seyferts that have simple Gaussian models to the Fe K$\alpha$ line, there are increasingly longer X-ray observations which allow physically motivated modeling of the reflection spectra. \cite{Walton12} have 16 Seyferts with with extensive reflection model fits to the X-ray spectra, and we include an additional 5 Seyferts with reflection line modeling from the literature. We note that we compare the absolute value of the spin parameter, $a$, to the Doppler-corrected radio luminosity. However, this is only important in one source, 3C 120, which has a negative spin \citep{Cowperthwaite12}.

In addition to the Seyferts, many stellar-mass black hole spectra have broad Fe line emission with enough signal-to-noise to warrant reflection line modeling. We proceed to use these as direct comparison to the Seyferts. Finally, we include neutron stars, which have spin estimates using coherent X-ray pulsations mission, burst oscillations and the difference between quasi-periodic oscillations. Although these are different techniques, modeling of the fluorescent Fe line in neutron stars does not allow us to probe the spin of these objects, as the radius of a neutron star is well beyond the deterministic range of 1.25--9$R_G$. These objects do probe lower spin and, and may still be informative in this study. We also note that the radio luminosity in these sources comes from a number of different accreting states, similar to the BHB sample, and are not corrected for Doppler boosting, unlike all of our black hole samples. We stress that the neutron stars are only meant to be illustrative.

Similar to the Gaussian line widths, we find that our samples shows little evidence of a correlation with the measured  Doppler-corrected radio luminosity and spin measurements (See Table \ref{tab:corr}). Figure \ref{fig:avr} plots Doppler-corrected radio luminosity versus spin, and Figure \ref{fig:zva} shows the spin dependence on redshift. However, when we plot radio luminosity per unit mass versus spin, we find that each of our black hole samples show a tentative positive correlation, in contrast to the neutron stars (See Figure \ref{fig:avpjet} \& Table \ref{tab:corr}). Further, when the entire compact object sample across the mass scale is considered, a positive correlation at 3.3$\sigma$ confidence level is found using the partial correlation test. Again, we point out that the neutron stars are not Doppler boosted corrected, which can introduce a range of correction factors from a factor of $~$0.002 at $\theta=0^\circ$ to a factor of $~80$ at $\theta=90^\circ$. This is assuming $\Gamma =5$, $k=3$ and $\alpha = -0.3$. We therefore stress that this correlation is only tentative, as we do not have the inclination nor the spectral index of the neutron star emission. However, if the correlation is true, the positive correlation we do find may point to spin being a common driving mechanism of jet luminosity in compact objects.

In Figure  \ref{fig:avm2} we plot the radio luminosity per unit mass squared versus spin. We note that the each black hole sample shows a tentative positive correlation, but the neutron stars and the entire sample as a whole, show a tentative negative correlation. Clearly, in Figure \ref{fig:avm2}, there is a segregation of the low mass sources from the high  mass sources, and combining them into one correlation is not likely to be informative as it reflects the mass segregation and not necessarily the dependence on spin.
 
We next consider the effects of spin when the mass accretion rate of the compact object is divided out. Figure \ref{fig:avrb} shows $\nu L_{\nu,5GHz} /L_{Bol}$ versus spin, where we have assumed that the Bolometric luminosity is a proxy for the total mass accretion rate as a general treatment, and do not take into account the accretion efficiency of individual systems. Shown Table \ref{tab:corr}, there is tentative evidence in the Seyferts, as well as the black hole binaries, that $\nu L_{\nu,5GHz} /L_{Bol}$ correlates with spin. When including neutron stars in the overall sample, the positive correlation is further strengthened to the 4.4$\sigma$ confidence level. This is suggesting that the radio luminosity per Bolometric luminosity is significantly correlated with the spin of the compact object. We extend this one step further and include the effects of both mass and mass accretion rate by dividing the Doppler-corrected radio luminosity by $L_{Bol}^{0.67}/M_{BH}^{0.78}$. This expression is derived from the fundamental plane of accretion onto Black Holes \citep{Gultekin09}, which links the mass, mass accretion rate and jet power across the mass scale in black holes. See Figure \ref{fig:avfund}. We find a tentative positive correlation in the black hole sample, which is further strengthened to 4.1$\sigma$ confidence when including the neutron stars. Although only suggestive, when taking out the effects of both mass and mass accretion rate, spin still has a role to play in determining the jet luminosity.

\section{Discussion}
\subsection{Gaussian Line Widths}
In all of the trends of our jet power proxies versus line width, we find little evidence of a correlation. If we accept the tentative inverse correlation, it may suggest that the jet power produced by an accreting black hole is inversely correlated with spin. There are a few theoretical predictions, which state that the most power jets may be produced by retrograde spins because they have the largest change in momentum between the accreting matter and the spin of the black hole \citep[e.g.,][]{Garofalo10}.  

However, the correlation is weak and may actually be suggesting that the line width from a Gaussian fit may not be a sufficient way to diagnose the properties of the emitting region, as we do not expect the line width to be symmetric and accurately described by a Gaussian line. Moreover, the statistically stronger correlation between X-ray and radio luminosity suggests that mass accretion rate does have a role in determining the amount of jet production, especially in comparison with the role of spin.

\subsection{Spin Measurements from Compact Objects}
When using more realistic models for the Fe K$\alpha$ line profiles, we find that Doppler-corrected radio luminosity, radio luminosity per unit mass, and radio luminosity per unit mass squared tentatively correlate with spin measurements of black holes . In addition, we find a suggestion of a correlation between spin measurements and Doppler-corrected radio luminosity per Bolometric luminosity, whether simply as Bolometric luminosity or in conjunction with mass as determined by the fundamental plane (i.e., $\log  (\nu L_R/L_{Bol}^{0.67}/M_{BH}^{0.78})$). This indicates spin may have a role to play in jet luminosity that is produced in these black holes, contrary to what was tentatively suggested by the Gaussian line width results. As mentioned in the previous subsection, the Gaussian line widths may not be an accurate measure of the fluorescent line profile. In addition, inclination may dominate over the spin as the primary driver of the line width and the previous result may be a depicting this effect. 

When we include neutron stars, although they show a negative correlation in each of our radio luminosity per $M_{BH}^{0}$, $M_{BH}^{1}$, \& $M_{BH}^{2}$, as part of the compact object ensemble, the sources show a statistically significant positive correlation between radio luminosity per unit mass as well as radio luminosity per Bolometric luminosity and spin, at over a 3.3$\sigma$ confidence level each. We do not expect neutron stars to tap spin via processes like \cite{Blandford77}. However, one could expect that the neutron star can be spun down to power jets. In fact, these positive correlations may imply that {\it all} compact objects are able to extract angular momentum to power jets in a similar manner. As evidenced in Figure \ref{fig:avrbbhb}, the average spin of the neutron stars is approximately 0.15 and the average spin of the BHB is approximately 0.9. The magnitude difference of radio flux one might expect from a \cite{Blandford77} jet model would be 36. This is the magnitude difference of jet luminosity \cite{Migliari03} find for their sample of neutron stars and BHB.  Further, the strong correlation between spin and radio luminosity per Bolometric luminosity demonstrates that not only is spin important in determining the jet luminosity, but the mass accretion rate is as well. 

This is also suggested in Figures \ref{fig:avpjet} \& \ref{fig:avm2}. The sources appear to be bounded by an upper limit while filling the parameter space below the dashed line ($\propto a^2$). Further more, the overall segregation and magnitude difference between the stellar-mass and supermassive black holes in Figure \ref{fig:avm2} can roughly be ascribed to the magnetic field strength. This is assuming the magnetic field energy density is in equipartition with the radiation energy density, i.e., $B^2/8\pi \simeq L_{Bol}/2\pi R^2$. Together, the apparent upper bound and magnitude offset of the samples in Figure \ref{fig:avm2} by mass, may imply that the spin of the black hole can set the maximum amount of power that the jet can extract from the system, while parameters like mass accretion rate and disk and/or coronal magnetic field strength produce the observed scatter under this bound. This suggests that all of these parameters interact together to produce the observed radio luminosity. The mass accretion rate and magnetic field would be analogous to a ``throttle", driving the radio luminosity vertically in Figures \ref{fig:avpjet} and \ref{fig:avm2}, which is ultimately set by the Eddington limit and the spin of the compact object. 

There are a few other works that have also looked at the effects of mass accretion and spin on jet power. \cite{Fender10} examine the role of jet power in X-ray binaries by analyzing sources in a low mass accretion state with steady, compact jet emission. They find no evidence for black hole spin powering jets from X-ray binaries. In addition, \cite{Russell13} found a similar lack of a correlation when examining transient jets from highly accreting X-ray binaries. By considering the different accreting regimes separately, each study was trying to isolate the effects of spin on jet production while keeping the effects of mass accretion rate constant. In addition, \cite{Fender10} note that there was either no correlation between spin and jet power, or that the data may be in error either because jet luminosity measurements and/or spin constraints are faulty.  The limited sample size examined in that work would make the results especially prone to even just a subset of the data being inaccurate in some manner, and this emphasized the need to examine a larger sample of sources.  In our work, then, we have directly examined the effects of mass accretion rate on different proxies for total jet power. We explicitly included the influence of the mass accretion rate by dividing the radio luminosity by the X-ray luminosity, our accretion rate proxy.  In that we have explored a number of different jet power proxies, a larger range in mass accretion rate, different proxies for the mass accretion rate, a large sample of sources, different angles on black hole spin, and a much larger range in black hole mass, the results of \cite{Fender10} and \cite{Russell13} might be regarded as special or limiting cases of our analysis, rather than discrepant results. 

Examining Figure \ref{fig:avpjet}, it is interesting to note that \cite{King13a} find a similar result, in that when comparing X-ray wind power across the black hole mass scale, the AGN were systematically at a higher wind power per unit mass when compared to their stellar-mass counter parts, consistent with the magnitude difference in Figure \ref{fig:avpjet}.  Although, we do not expect the X-ray winds to be driven by the \cite{Blandford77} mechanism, magnetic fields could play a key role in launching and driving these winds and regulate the power of the outflows, albeit winds or jets, \citep[e.g.,][]{lovelace76,Blandford82}.

\subsection{Spin contribution to Observational Trends: Radio Loud and Quiet Dichotomy}

There have been suggestions that the difference between radio-loud and radio-quiet galaxies is a result of the difference in spin in the black hole \citep[e.g.,][]{Moderski98}. We define radio-loud as having $R\gtrsim$10, where $R$ is the ratio between the jet luminosity at 5 GHz and the optical luminosity at 4400\AA\/ (B band). This scenario postulates that the radio-loud black holes have high spin, while the radio-quiet have low spin \citep[e.g.,][]{Wilson95}. This assumes that more energy can be tapped from high spin black holes, thus producing stronger jet emission. Our sample is predominantly radio-quiet with the exception of a few sources, and predominantly high spin, which contradicts this claim. This is also in conjunction with the findings of \cite{Walton12} 

There have been several claims that a retro-grade spin could power more powerful jets than pro-grade \citep{Garofalo10}. This would seem to agree with our Figure \ref{fig:sigvr} where the Fe K$\alpha$ line width inversely correlates with radio luminosity, if line width is positively correlated with spin. However, the line width also is likely to correlate with inclination and emissivity of the emitting regions. Additionally, the mass of many AGN is not well constrained and may play the dominant role in this radio-loud/radio-quiet dichotomy \citep[e.g.,][]{Broderick11}. Environment, mass, and mass accretion rate have all been suggested as primary factors to this dichotomy over spin \citep[e.g.,][]{Broderick11}. In addition, radio-loud galaxies may still need high spin, but radio-quiet does not necessarily imply low spin.

In addition, recent work by \cite{Sikora13} suggests that the radio-loud versus radio-quiet dichotomy may be a result of ordered magnetic field strength and magnetically arrested/chocked accretion flows. They suggest that radio-loud quasars have stronger magnetic fields and are capable of producing larger amounts of radio luminosity than their radio-quiet counter parts. We also note that \cite{King13b} find that neutron stars are also consistent with this paradigm of magnetically arrested accretion flows. However, this magnetic flux paradigm and jet dependence on spin may not be mutually exclusive. 

\subsection{Assumptions, Caveats, and Biases}
\label{Bias}
In this study we have made a number of assumptions about the observables and how they relate to the outflows being studied. We have assumed that the compact radio emission is a direct proxy for the jet power, and we note that there is scatter in the \cite{Merloni07} relation between radio luminosity and jet power. In addition, there may also be unresolved features in the radio observations, such as impact lobes analogous to FR I galaxies, that could contaminate our measurements, making the radio luminosities used here, effective upper limits. We have also neglected the role of environment in determining the radio luminosity. One can imagine that in denser environments, there would be more shocks and thus stronger radio emission \citep[e.g.,][]{Bicknell95}. There may even be a systematic bias toward larger black holes with larger radio luminosity, for if the larger black holes preferentially reside in groups and clusters, their environment may be denser and produce larger radio emission in consequent shocks. 

In addition, in Figures \ref{fig:sigvm2} and \ref{fig:avm2}, we have made the assumption that the \cite{Blandford77} model is correct and that the radio luminosity is correlated with the energy carried away by the jet. In the \cite{Blandford77} model, the jet is generated from the poynting flux of the magnetic fields, while the radio emission is generated further down the outflow as non-thermal emission via synchrotron emission. The \cite{Blandford77} model also makes the assumption that the accretion disk luminosity can be compared to jet luminosity, if it is ``electromagnetic". However, at low accretion regimes, i.e. $10^{-3} L_{Edd}$, the accretion disk may not be a geometrically thin, optically thick Shakura-Sunyaev accretion disk \citep{Shakura73}, but a less efficient emitter, like an advective dominated accretion flow (ADAF). BHB typically emit compact radio emission in this regime, in the ``low/hard" state, but the majority of radio detections in our study were taken during radio outbursts, typically in the transition between X-ray states. Therefore, this may not be a problem in our black hole sample. This is also pertinent for the the neutron star sample, which includes several different X-ray states (Atoll, Z-types, and accreting millisecond X-ray pulsars). However, as this study is preliminary, we seek to examine the state of the field as it stands, not to exclude sources based on personal biases.

Examining the X-ray emission further, we note that in samples of Gaussian line fits, both \cite{Patrick12} include partial covering in their initial fits to the X-ray continuum. Variability studies do not appear to support partial covering \citep[e.g.,][]{Turner04}. The inclusion of absorption in the spectra may serve to bias their line widths, reducing the contribution of the red wing of the reflection lines. This would serve to narrow the line fits. However, it is unclear if this systematic bias to lower line widths would introduce the slight inverse correlation that is observed. \cite{Crenshaw12} have shown a positive correlation with column density of warm-absorbers and X-ray luminosity. As Figure \ref{fig:xvr} shows a positive correlation with X-ray luminosity and radio luminosity, we could expect a positive correlation between warm-absorbers and radio luminosity as well. If this is the case, partial covering to model warm absorbers could be more prevalent in more luminous radio sources and thus influencing the line widths to be narrower. This could potentially bias the data, if excessively added, and produce the slight inverse correlation we observe in Figure \ref{fig:sigvr}, \ref{fig:sigvpjet}, \& \ref{fig:sigvm2}. 

 An additional effect of detecting lines in low signal-to-noise should also be considered before trusting the inverse correlation between line width and radio luminosity. At low signal-to-noise, a broad line may not stand out above the continuum. Therefore a line of a given equivalent width may only be detected if it is narrow as compared to a broad feature that blends into the continuum. As the sample is not flux limited, we may be subject to this kind of bias.

It is important to note that the Seyfert radio measurements are not simultaneous with the X-ray measurements, unlike the BHB and neutron star samples. The Seyfert X-ray flux is known to be highly variable, while the radio shows less fluctuations \citep[e.g.,][]{King11}. This can conceivably introduce scatter into our analysis when employing both the radio and Bolometric luminosity (See Figures \ref{fig:sigvrb}, \ref{fig:avrb}, \& \ref{fig:avrbbhb}). In addition, the Bolometric correction used in our analysis is assumed to be 20 for the majority of the galaxies \citep{Vasudevan09}. This could be off by at least an order of magnitude, adding an additional source of scatter to the relation. These two effects could be the reason that BHB and neutron stars show a tighter relation in Figure \ref{fig:avrbbhb} than the Seyferts do in Figure \ref{fig:avrb}.

Finally, including neutron stars in the sample may introduce a bias. While black holes, in theory, can traverse the entire range of spin, from retrograde to maximally spin prograde, neutron stars can only reach a spin of $-0.8\lesssim a \lesssim0.8$ before breaking up.  This limitation of neutron stars may suggest the two types of compact objects follow different relations in reference to spin. In addition, the presence of a surface in a neutron star may also alter the way spin is tapped, as compared to black holes. However, if its the change in spin that is essential to power jets, than the two types of compact objects can justifiably be compared. Additionally, the precision of the spin in neutron stars vastly exceeds that of the black hole spin measurements, and is a limitation of this work. 

\section{Conclusions}
In this paper, we have attempted to compare the spin and jet power in compact objects. We find that:
\begin{itemize}
\item Seyfert Fe K$\alpha$ lines modeled with Gaussian lines show a slight inverse correlation to Doppler-corrected radio luminosity and jet power proxies.
\item There is evidence of a tentative positive correlation between reflection spin measurements and our jet power proxies in the black hole samples (i.e., Seyferts and stellar-mass black holes).
\item When including neutron stars (to extend the range in spin parameters) the positive correlation between spin and our jet power proxies becomes slightly more statistically significant. 
\item Our study has tentative evidence that spin is important in determining the jet power in compact accreting systems. More sources, in a wider range of spins, are needed to better determine this correlation.
\item The mass accretion rate and disk and/or coronal magnetic field strength may be analogous to a ``throttle" while mass and spin may set the maximum amount of jet power in these accreting systems, as evidenced by Figures \ref{fig:avpjet} and \ref{fig:avm2}.
\item We note that the majority of sources in our AGN sample are radio-quiet Seyferts that have high spin. This contradicts the idea that the radio-loud vs radio-quiet dichotomy is driven by spin, where high spin generates the radio-loud sample and low spin generates the radio-quiet sample.  
\item Finally, we note  that our results may be subject to biases that could arise from an incomplete sample, modeling of X-ray absorption features, non-simultaneous X-ray and radio measurements, particular jet models, and comparison of different compact objects. Now multi-wavelength observations across the mass scale are necessary in order to make additional progress.  
\end{itemize}
  
\section*{Acknowledgements}
We would like to thank the anonymous referee for their invaluable report. ALK acknowledges support from NASA through the NESSF program. JMM thanks NASA for support through its guest observer programs.

\clearpage

\LongTables
\begin{deluxetable*}{l l l l l l l l l l  }[h]
\tablecolumns{10}
\tablewidth{0pc}
\tabletypesize{\scriptsize}
\tablecaption{Data Parameters}
\tablehead{ Source & $\log L_{X-ray}$ & $\log \nu L_\nu(5GHz)$ &  $\sigma$ & $a$  & $\theta$ & $\alpha$ &$\log M_{BH}$ &  z & Ref\\  }
\startdata
\multicolumn{3}{l}{\bf \cite{Nandra07}}\\
Ark 120 &  43.89$\pm  0.04$ &  41.45$^{+  0.28}_{-  0.58}$ &   0.20$^{+  0.08}_{-  0.06}$ &-& 82$^{+  3}_{- 30}$ &  -0.70$\pm$  0.30 &   8.27 &    0.033 & 1,2,3\\
IC 4329A &  43.66$\pm  0.04$ &  39.70$^{+  0.33}_{- 10.53}$ &   0.40$^{+  0.09}_{-  0.07}$ &-& 60$^{+ 25}_{- 60}$ &  -0.70$\pm$  0.30 &   6.69 &    0.016 & 4\\
MCG -05-23-16 &  42.94$\pm  0.04$ &  33.29$^{+  0.38}_{-  0.21}$ &   0.55$^{+  0.12}_{-  0.10}$ &-&  0$^{+ 19}_{-  0}$ &  -1.07$\pm$  0.05 &   6.30 &    0.008 & 5,26 \\
MCG -06-30-15 &  42.71$\pm  0.04$ &  35.26$^{+  0.27}_{-  1.29}$ &   0.69$^{+  0.10}_{-  0.11}$ &-& 20$^{+  4}_{- 20}$ &  -0.70$\pm$  0.30 &   6.46 &    0.008 & 5,11\\
Mrk 766 &  42.89$\pm  0.04$ &  38.75$^{+  0.22}_{-  0.25}$ &   0.53$^{+  0.19}_{-  0.11}$ &-& 40$^{+  5}_{-  6}$ &  -0.70$\pm$  0.30 &   6.64 &    0.013  & 1,4\\
NGC 2992 &  42.99$\pm  0.04$ &  38.10$^{+  0.26}_{-  0.36}$ &   0.32$^{+  0.25}_{-  0.15}$ &-& 24$^{+  7}_{-  7}$ &  -0.70$\pm$  0.30 &   7.75 &    0.008 & 1,4 \\
NGC 3516 &  42.37$\pm  0.04$ &  37.25$^{+  0.21}_{-  0.24}$ &   0.51$^{+  0.11}_{-  0.09}$ &-& 31$^{+  2}_{-  4}$ &  -1.30$\pm$  0.30 &   7.50 &    0.009 &4,6 \\
NGC 3783 &  42.97$\pm  0.04$ &  34.14$^{+  0.39}_{-  0.22}$ &   0.84$^{+  0.47}_{-  0.08}$ &-&  0$^{+ 19}_{-  0}$ &  -0.97$\pm$  0.09 &   7.47 &    0.010 & 7,8\\
NGC 4051 &  41.25$\pm  0.04$ &  34.84$^{+  0.27}_{-  1.04}$ &   0.17$^{+  0.01}_{-  0.30}$ &-& 22$^{+  6}_{- 15}$ &  -0.68$\pm$  0.30 &   6.23 &    0.002 & 4,6 \\
NGC 4151 &  42.59$\pm  0.04$ &  36.25$^{+  0.35}_{-  0.95}$ &   0.31$^{+  0.14}_{-  0.11}$ &-& 17$^{+ 12}_{- 17}$ &  -0.70$\pm$  0.30 &   7.12 &    0.003 & 14,8\\
NGC 4593 &  42.80$\pm  0.04$ &  36.59$^{+  0.44}_{-  1.24}$ &   0.50$^{+  0.26}_{-  0.15}$ &-& 24$^{+ 61}_{- 17}$ &  -0.11$\pm$  0.30 &   6.99 &    0.009 & 9,10\\
NGC 526A &  43.18$\pm  0.04$ &  38.79$^{+  0.37}_{-  0.85}$ &   0.12$^{+  0.19}_{-  0.13}$ &-& 43$^{+ 42}_{- 20}$ &  -0.70$\pm$  0.30 &   - &    0.019 & 1\\
NGC 5506 &  42.74$\pm  0.04$ &  40.29$^{+  0.31}_{-  0.50}$ &   0.32$^{+  0.10}_{-  0.07}$ &-& 58$^{+ 27}_{- 19}$ &  -0.31$\pm$  0.30 &   7.94 &    0.007 & 9,12\\
NGC 5548 &  43.37$\pm  0.04$ &  36.39$^{+  0.49}_{-  0.76}$ &   0.71$^{+  0.08}_{-  0.06}$ &-& 15$^{+ 70}_{- 15}$ &  -0.70$\pm$  0.30 &   7.65 &    0.017 & 4,6\\
NGC 7314 &  42.28$\pm  0.04$ &  37.97$^{+  0.21}_{-  0.22}$ &   0.87$^{+  0.17}_{-  0.08}$ &-& 42$^{+  3}_{-  4}$ &  -0.70$\pm$  0.30 &   $<$9.67 &    0.005 & 1,13 \\
\multicolumn{3}{l}{\bf \cite{Patrick12}}\\

3C 120 &  43.93$\pm  0.04$ &  40.26$^{+  0.26}_{-  0.26}$ &   0.12$^{+  0.03}_{-  0.02}$ &  - &  17$\pm$  1 &   0.09$\pm$  0.30 &   7.36 &    0.033 & 4\\
3C 382 &  44.40$\pm  0.04$ &  41.44$^{+  0.22}_{-  0.22}$ &   0.20$^{+  0.21}_{-  0.09}$ &  - &  30$\pm$  3 &  -0.40$\pm$  0.30 &   9.06 &    0.058 & 14,16\\
3C 390.3 &  44.26$\pm  0.04$ &  41.94$^{+  0.22}_{-  0.22}$ &   0.32$^{+  0.26}_{-  0.11}$ &  -&  49$\pm$  3 &  -0.70$\pm$  0.30 &   8.46 &    0.056 & 4,8\\
IC 4329A &  43.72$\pm  0.04$ &  39.96$^{+  0.24}_{-  0.24}$ &   0.55$^{+  0.12}_{-  0.10}$ &  - &  51$\pm$  4 &  -0.35$\pm$  0.14 &   6.69 &    0.016 & 4\\
MCG -05-23-16 &  43.09$\pm  0.04$ &  36.25$^{+  0.22}_{-  0.22}$ &   0.51$^{+  0.09}_{-  0.08}$ &   $<$0.50 &  24$\pm$  3 &  -1.07$\pm$  0.05 &   6.30 &    0.009 & 5,26\\
MCG -06-30-15 &  42.67$\pm  0.04$ &  37.34$^{+  0.23}_{-  0.23}$ &   0.84$^{+  0.06}_{-  0.06}$ &   0.49 $^{+  0.20}_{-  0.10 }$ &  44$\pm$  4 &  -0.70$\pm$  0.30 &   6.46 &    0.008& 5,11 \\
MCG +8-11-11 &  43.71$\pm  0.04$ &  37.79$^{+  0.26}_{-  0.26}$ &   0.17$^{+  0.06}_{-  0.04}$ &  - &  18$\pm$  2 &  -0.70$\pm$  0.30 & - &    0.021 &  14\\
MR 2251-178 &  44.51$\pm  0.04$ &  39.85$^{+  0.25}_{-  0.25}$ &   0.31$^{+  0.20}_{-  0.10}$ &  - &  36$\pm$  7 &  -0.70$\pm$  0.30 &   - &    0.064 & 1\\
Mrk 79 &  43.14$\pm  0.04$ &  38.88$^{+  0.21}_{-  0.21}$ &   0.53$^{+  0.21}_{-  0.16}$ &   $<$0.80  &  34$\pm$  3 &  -0.70$\pm$  0.30 &   - &    0.022 & 1\\
Mrk 335 &  43.27$\pm  0.04$ &  38.50$^{+  0.21}_{-  0.21}$ &   0.50$^{+  0.13}_{-  0.11}$ &   0.70 $^{+  0.12}_{-  0.01 }$ &  38$\pm$  2 &  -0.29$\pm$  0.26 &   7.15 &    0.026 & 4,8\\
Mrk 509 &  44.03$\pm  0.04$ &  38.71$^{+  0.22}_{-  0.22}$ &   0.69$^{+  1.36}_{-  0.15}$ &  - &  41$\pm$  4 &  -0.56$\pm$  0.30 &   8.16 &    0.034 & 4,8\\
Mrk 841 &  43.55$\pm  0.04$ &  38.63$^{+  0.44}_{-  0.25}$ &   0.40$^{+  0.29}_{-  0.17}$ &  - &  $>$32 &  -0.74$\pm$  0.30 &   7.88 &    0.036 &17,18 \\
NGC 2992 &  42.12$\pm  0.04$ &  38.28$^{+  0.44}_{-  0.21}$ &   0.32$^{+  0.15}_{-  0.10}$ &  - &  $>$26 &  -0.70$\pm$  0.30 &   7.75 &    0.008 &1,4\\
NGC 3227 &  42.05$\pm  0.04$ &  37.27$^{+  0.21}_{-  0.21}$ &   0.71$^{+  0.10}_{-  0.09}$ &  - &  33$\pm$  2 &  -0.90$\pm$  0.30 &   6.88 &    0.004 & 4,6 \\
NGC 3516 &  42.54$\pm  0.04$ &  38.14$^{+  0.21}_{-  5.17}$ &   0.87$^{+  0.12}_{-  0.10}$ &   $<$0.30  &  $<$41 &  -1.30$\pm$  0.30 &   7.50 &    0.009 & 4,6\\
NGC 3783 &  42.91$\pm  0.04$ &  36.84$^{+  0.20}_{-  1.69}$ &   0.76$^{+  0.22}_{-  0.10}$ &   $<$0.24  &  $<$23 &  -0.97$\pm$  0.09 &   7.47 &    0.010 & 7,8\\
NGC 4051 &  40.94$\pm  0.04$ &  34.73$^{+  0.24}_{-  1.29}$ &   0.74$^{+  0.15}_{-  0.12}$ &  - &  $<$20 &  -0.68$\pm$  0.30 &   6.23 &    0.002 & 4,6\\
NGC 5506 &  42.87$\pm  0.04$ &  37.59$^{+  0.27}_{-  0.27}$ &   0.32$^{+  0.07}_{-  0.06}$ &  - &  20$\pm$  4 &  -0.31$\pm$  0.30 &   7.94 &    0.006 &9,12\\
NGC 7469 &  43.02$\pm  0.04$ &  38.22$^{+  0.39}_{-  0.39}$ &   0.15$^{+  0.07}_{-  0.03}$ &   0.69 $\pm$ 0.09  &  23$\pm$ 11 &  -0.59$\pm$  0.16 &   7.09 &    0.016 & 4,8\\
\multicolumn{3}{l}{\bf \cite{Walton12}+5 AGN}\\
1H 0323+342 &  43.76$\pm  0.04$ &  41.71$\pm{  0.21}$ & - &   $>$0.48  &  45* &  -0.30$\pm$  0.30 & - &    0.061 & 14\\
**3C 120 &  43.92$\pm  0.04$ &  40.38$\pm{ 0.27}$ & - &  $<$-0.10  &  18.20$\pm$  3.10 &   0.09$\pm$  0.30 &   7.36 &    0.033 &  50,4\\
3C 382 &  44.40$\pm  0.04$ &  42.17$\pm{ 0.21}$ & - &   0.75 $^{+  0.07}_{-  0.04 }$ &  40* &  -0.40$\pm$  0.30 &   9.06 &    0.058 & 14,16\\
Ark 120 &  43.70$\pm  0.04$ &  40.30$\pm{ 0.24}$ & - &   0.81 $^{+  0.10}_{-  0.18 }$ &  54$\pm$  6 &  -0.70$\pm$  0.30 &   8.27 &    0.033 & 1,2,3 \\
Ark 564 &  43.46$\pm  0.04$ &  40.50$\pm{ 0.22}$ & - &   0.96 $^{+  0.01}_{-  0.06 }$ &  64$\pm$  6 &  -0.74$\pm$  0.10 &   6.50 &    0.025 & 4\\
Fairall 9 &  43.91$\pm  0.04$ &  $<$40.15 & - &   0.64 $^{+  0.00}_{-  0 }$ &  45$\pm$ 11 &  -0.70$\pm$  0.30 &   7.00 &    0.047 & 4\\
**IRAS 00521 &  43.02$\pm  0.04$ &  39.57$\pm{  0.22}$ & - &   0.97 $^{+  0.03}_{-  0.13 }$ &  37$\pm$  4 &  -0.70$\pm$  0.30 &   - &    0.069 & 48,49\\
IRAS 13224 &  42.51$\pm  0.04$ &  41.06$\pm{ 0.25}$ & - &   $>$0.99  &  65*&  -0.70$\pm$  0.30 &   6.75 &    0.066 & 15,25\\
**MCG -06-30-15 &  42.56$\pm  0.04$ &  36.96$\pm{  0.22}$ & - &   0.97 $\pm{  0.03 }$ &  38$\pm$  3 &  -0.70$\pm$  0.30 &   6.46 &    0.008 & 47,5,11\\
Mrk110 &  43.54$\pm  0.04$ &  38.70$\pm{  0.24}$ & - &   $>$0.99 &  31$\pm$  5 &  -0.70$\pm$  0.30 &   6.70 &    0.035 & 1,2,3\\
Mrk 335 &  43.19$\pm  0.04$ &  39.20$\pm{  0.25}$ & - &   0.83 $^{+  0.10}_{-  0.13 }$ &  50$\pm$  8 &  -0.29$\pm$  0.26 &   7.15 &    0.026 & 4,8\\
Mrk 359 &  42.39$\pm  0.04$ &  38.95$\pm{ 0.24}$ & - &   0.66 $^{+  0.30}_{-  0.46 }$ &  47$\pm$  6 &  -0.70$\pm$  0.30 &   6.33 &    0.017 & 23,24\\
Mrk 509 &  43.97$\pm  0.04$ &  36.64$^{+  0.25}_{-  1.06}$ & - &   0.86 $^{+  0.02}_{-  0.01 }$ &  $<$18 &  -0.56$\pm$  0.30 &   8.16 &    0.034 & 4,8\\
Mrk 841 &  43.49$\pm  0.04$ &  39.60$\pm{  0.27}$ & - &   $>$0.56 &  45$\pm$  6 &  -0.74$\pm$  0.30 &   7.88 &    0.036 & 17,18\\
Mrk 1018 &  43.49$\pm  0.04$ &  39.54$\pm{  0.29}$ & - &   0.57 $^{+  0.31}_{-  0.82 }$ &  45$\pm$ 12 &  -0.70$\pm$  0.30 &   - &   0.042 & 23 \\
**NGC 3783 &  42.98$\pm  0.04$ &  36.98$\pm$ 0.22 & - &   0.92 $^{+  0.07}_{-  0.03 }$ &  24$\pm$  3 &  -0.97$\pm$  0.09 &   7.47 &    0.010& 52,7,8 \\
NGC 7469 &  42.97$\pm  0.04$ &  40.47$^{+  0.21}_{-  8.68}$ & - &   $>$0.96  &  $<$54 &  -0.73$\pm$  0.14 &   7.09 &    0.016 &4,8\\
PDS 456 &  44.35$\pm  0.04$ &  42.38$\pm{  0.21}$ & - &   $>$0.97  &  70$\pm$  4 &  -0.39$\pm$  0.09 &   9.00 &    0.184 & 21,22\\
PKS 0558-504 &  44.59$\pm  0.04$ &  41.96$\pm{  0.20}$ & - &   $>$0.80&  45* &  -0.84$\pm$  0.16 &   8.70 &    0.137 & 19,20\\
Swift J0501.0-3239 &  42.81$\pm  0.04$ &  38.61$^{+  0.22}_{-  6.97}$ & - &   $>$0.96  &  $<$48 &  -0.70$\pm$  0.30 & - &    0.012 & 1\\
**Swift J2127.4+5654 &  43.14$\pm  0.04$ &  38.85$\pm{  0.22}$ & - &   0.60 $\pm{  0.20 }$ &  46$\pm$  4 &  -0.70$\pm$  0.30 &   7.20 &    0.014 & 1,51 \\

{\bf BHB} & & & &&& & & {\bf d(kpc)} \\
4U 1543-475 &  35.91$\pm  0.11$ &  29.19$\pm$  0.11 & - &   0.30 $\pm{  0.10 }$ &  22$\pm$  1 &   0.08$\pm$    0.05 & 
  0.95 &   8 & 27,33\\
Cyg X-1 &  36.68$\pm  0.04$ &  29.98$\pm$  0.23 & - &   0.97 $^{+  0.01}_{-  0.02 }$ &  24$\pm$  6 &  -0.30$\pm$    0.39 & 
  1.17 &   2.50 & 31,42\\
GRO J1655-40 &  35.32$\pm  0.14$ &  30.83$\pm$  0.16 & - &   0.98 $\pm{  0.01 }$ &  69$\pm$  1 &   0.10$\pm$    0.15 & 
  0.85 &   3.20 & 27,36,44 \\
GRS 1915+105 &  38.06$\pm  0.13$ &  34.01$\pm$  0.10 & - &   0.98 $\pm{  0.01 }$ &  55$\pm$  2 &  -0.54$\pm$    0.17 & 
  1.15 &  11 & 46,43,32\\
GX 339-4 &  37.81$\pm  0.23$ &  30.86$\pm$  0.23 & - &   0.94 $\pm{ 0.02 }$ &  29$\pm$  2 &   0.27$\pm$    0.01 & 
  0.78 &  11.50 & 27,37 \\
MAXI J1836-194 &  36.93$\pm  0.33$ &  29.90$^{+  0.33}_{-  0.98}$ & - &   0.88 $\pm{ 0.03 }$ &  $<$17&   0.72$\pm$    0.01 & 
  1* &   8* & 30,41\\
XTE J1550-564 &  38.42$\pm  0.12$ &  32.59$\pm$  0.08 & - &   0.76 $\pm{ 0.01 }$ &  50$\pm$  1 &  -0.32$\pm$    0.02 & 
  1.04 &   5.50 & 27,34\\
XTE J1650-500 &  37.92$\pm  0.33$ &  30.91$\pm$  0.33 & - &   0.79 $\pm{  0.01 }$ &  45$\pm$  1 &  -0.28$\pm$    0.14 & 
  0.70 &   8 & 27,35\\
XTE J1652-453 &  37.28$\pm  0.33$ &  26.80$\pm$  0.38 & - &   0.45 $\pm{  0.02 }$ &   8.8$\pm$  0.1 &  -0.47$\pm$    0.23 & 
  1* &   8* & 28,39,45\\
XTE J1752-223 &   -&  29.47$\pm$  0.37 & - &   0.52 $\pm{  0.11 }$ &  28$\pm$  6 &  -0.30$\pm$    0.39 & 
  1* &   8* & 29,40\\
XTE J1908+094 &  - &  30.27$\pm$  0.36 & - &   0.75 $\pm{ 0.09 }$ &  45$\pm$  8 &  -0.49$\pm$    0.34 & 
  1* &   8* & 27,38\\
{\bf NS} \\
4U 0614-09 &  35.92$\pm  0.10$ &  $<$27.48& - &    0.169 $\pm$   0.002 & - & - &   0.15 &   $<$3.00 & 53,54\\
4U 1728-34 &  36.66$\pm  0.09$ &  28.69$\pm$  0.11 & - &    0.148 $\pm$   0.002 & - & - &   0.15 &   4.60 & 53,54\\
4U 1820-30 &  37.78$\pm  0.10$ &  28.64$\pm$  0.10 & - &    0.116 $\pm$   0.026 & - & - &   0.15 &   7.60 & 53,54\\
Aql X-1 &  36.51$\pm  0.10$ &  28.34$\pm$  0.11 & - &    0.224 $\pm$   0.000 & - & - &   0.15 &   5.20 & 53,54\\
Cyg X-2 &  38.36$\pm  0.10$ &  29.60$\pm$  0.17 & - &    0.143 $\pm$   0.014 & - & - &   0.15 &  13.30 & 53,54 \\
GX 17+2 &  38.48$\pm  0.10$ &  29.87$\pm$  0.16 & - &    0.111 $\pm$   0.020 & - & - &   0.15 &  14.00 & 53,54\\
GX 340+0 &  38.09$\pm  0.10$ &  29.43$\pm$  0.23 & - &    0.140 $\pm$   0.003 & - & - &   0.15 &  11.00 & 53,54\\
GX 349+2 &  37.63$\pm  0.10$ &  28.75$\pm$  0.23 & - &    0.108 $\pm$   0.005 & - & - &   0.15 &   5.00 & 53,54\\
GX 5-1 &  38.31$\pm  0.10$ &  29.61$\pm$  0.13 & - &    0.117 $\pm$   0.028 & - & - &   0.15 &   9.20 & 53,54\\
IGR J00291 &  35.82$\pm  0.10$ &  $<$28.70 & - &    0.163 $\pm$   0.000 & - & - &   0.15 &   $<$3.00 & 53,54\\
MXB 1730-335 &  37.43$\pm  0.10$ &  29.03$\pm$  0.09 & - &    0.125 $\pm$   0.002 & - & - &   0.15 &   8.80 & 53,54\\
SAX J1808.4 &  35.02$\pm  0.10$ &  28.27$\pm$  0.13 & - &    0.163 $\pm$   0.000 & - & - &   0.15 &   2.50 & 53,54\\
Sco X-1 &  38.38$\pm  0.10$ &  29.47$\pm$  0.16 & - &    0.111 $\pm$   0.016 & - & - &   0.15 &   2.80 & 53,54\\
\enddata
\label{tab:quant}
\tablecomments{All the Radio luminosities are Doppler-corrected except for the Neutron Stars.  $1)$ \cite{Condon98}, $2)$ \cite{Ho02}, $3)$ \cite{Kaspi00}, $4)$ \cite{Merloni03}, $5)$ \cite{Mundell09}, $6)$ \cite{Denney10}, $7)$ \cite{Ulvestad84}, $8)$ \cite{Peterson04}, $9)$ \cite{Gallimore06}, $10)$ \cite{Denney07},  $11)$ \cite{McHardy05}, $12)$ \cite{Nikolajuk09}, $13)$ \cite{Yaqoob96}, $14)$ \cite{Becker91}, $15) $\cite{Wang04} , $16)$ \cite{Marchesini04}, $17)$\cite{Edelson87},$18)$ \cite{Wandel02}, $19)$ \cite{Murphy10}, $20)$ \cite{Wang01}, $21)$ \cite{Yun04}, $22)$ \cite{Reeves00}, $23)$ \cite{Condon02}, $24)$ \cite{Hao05}, $25)$ \cite{Feain09}, $26)$ \cite{Beckmann08} , $27)$ \cite{Miller09} and references therein, $28)$ \cite{Hiemstra11}, $29)$ \cite{Reis11}, $30)$ \cite{Reis12}, $31)$ \cite{Fabian12b}, $32)$ \cite{Blum09} and references therein, $33)$ \cite{Kalemci05}, $34)$ \cite{Hannikainen09}, $35)$ \cite{Corbel04}, $36)$ \cite{Rupen05}, $37)$ \cite{Corbel13}, $38)$ \cite{Rupen02}, $39)$ \cite{Calvelo09}, $40)$ \cite{Yang11}, $41)$ \cite{MillerJones11}, $42)$ \cite{Fender06}, $43)$ \cite{Rodriguez95}, $44)$ \cite{Migliari07}, $45)$ \cite{Markwardt09}, $46)$ \cite{Sazonov94}, $47)$\cite{Miniutti07} , $48)$ \cite{Mauch03}, $49)$ \cite{Tan12}, $50)$ \cite{Cowperthwaite12}, $51)$ \cite{Miniutti09}, $52)$ \cite{Reynolds12}, $53)$ \cite{Migliari06}, $54)$ \cite{Migliari11},  \& * the mass and distance of these BHB is unknown and assumed to be 10$M_\odot$ and 8 kpc respectively. We have added radio errors of 0.2 dex in quadrature to the AGN, following \cite{Ho01} (2). The Seyfert X-ray errors are assumed to be 10\%, i.e. $~$0.4dex. ** These are the additional 5 Seyferts that included in addition to the \cite{Walton12} sample. 
}
\end{deluxetable*}
\clearpage

\bibliography{bibwinds}

\end{document}